\documentclass[aps,superscriptaddress,twocolumn,twoside,nofootinbib,floatfix,prl,a4paper]{revtex4-2}

\usepackage[utf8]{inputenc}
\usepackage[T1]{fontenc}     
\usepackage[british]{babel}  

\usepackage[sc,osf]{mathpazo}
\usepackage{times}

\usepackage[table,dvipsnames]{xcolor}

\usepackage[scaled=0.86]{berasans}

\usepackage{makecell}
\usepackage{comment}
\usepackage{enumitem}
\usepackage{graphicx}
\usepackage[babel]{microtype}

\usepackage{amsmath,amssymb,amsthm,bm,amsfonts,mathrsfs,bbm}

\usepackage{xspace}
\usepackage{pgf,tikz}
\usepackage{multirow}
\usepackage{array}
\usepackage{bigstrut}
\usepackage{braket}
\usepackage{mathtools}
\usepackage{subcaption}
\usepackage{natbib}
\usepackage{centernot}

\usepackage{algorithm}
\usepackage{algorithmic}
\usepackage{booktabs}
\usepackage{tabularx}

\usepackage[colorlinks=true,citecolor=Magenta,urlcolor=Blue]{hyperref}

\makeatletter
\newcommand{\setword}[2]{%
  \phantomsection
  #1\def\@currentlabel{\unexpanded{#1}}\label{#2}%
}
\makeatother

\newcommand{\Tr}{\operatorname{Tr}}

\newcommand{\be}{\begin{equation}}
\newcommand{\ee}{\end{equation}}
\newcommand{\ba}{\begin{eqnarray}}
\newcommand{\ea}{\end{eqnarray}}

\def\>{\rangle}
\def\<{\langle}

\providecommand{\ket}[1]{| #1{\rangle}}

\begin{document}
\title{Maximizing Nonclassicality of Massive Objects via Quantum Zeno Effect}

\author{Debarshi Das}
\email{dasdebarshi90@gmail.com}
\affiliation{Department of Physics, Shiv Nadar Institution of Eminence, Gautam Buddha Nagar, Uttar Pradesh 201314, India}

\author{Pritam Roy}
\email{roy.pritamphy@gmail.com}
\affiliation{S. N. Bose National Centre for Basic Sciences, Block JD, Sector III, Salt Lake, Kolkata 700 106, India}

\author{Marko Toroš}
\affiliation{Faculty of Mathematics and Physics, University of Ljubljana, Jadranska 19, SI-1000 Ljubljana, Slovenia}

\author{Hendrik Ulbricht}
\affiliation{School of Physics and Astronomy, University of Southampton, Southampton SO17 1BJ, England, United Kingdom}

\author{Dipankar Home}
\affiliation{Raman Research Institute (RRI), C. V. Raman Avenue, Sadashivanagar, Bengaluru, Karnataka 560080, India }
\affiliation{Bose Institute, Kolkata 700 091, India}

\author{Sougato Bose}
\affiliation{Department of Physics and Astronomy, University College London, Gower Street, London WC1E 6BT, England, United Kingdom}

\begin{abstract}
For testing quantum mechanics in the macroscopic domain, a major challenge is to devise effective means for enhancing the observable nonclassical signatures despite the ubiquitous presence of environmental decoherence. Toward this goal, we invoke the Quantum Zeno Effect (QZE) for achieving a tunable amplification of an inherently nonclassical quantum disturbance induced by any measurement. Such an enhancement of otherwise small and decoherence-suppressed nonclassicality  can arise from the cumulative quantum disturbances generated by repetitive measurements, with the tunability of amplification controlled by the number of measurements. To evidence this, we formulate a testable loophole-free scheme using a massive oscillator, where the system preparation requires trapping and ground-state cooling of a massive object. The required measurements can be realized through a beam-splitter-type interaction between the mechanical oscillator and an optical field, followed by photon detection. 
Our analysis shows that such amplification, suitably quantified in terms of a testable witness, remains appreciably observable even in the realistic regimes of optomechanical damping, and for sufficiently large masses, thus enabling the demonstration of QZE in the macroscopic domain.
\end{abstract}

\maketitle

{\em Introduction.---}Testing and exploring ramifications of quantum mechanical principles at the macroscopic level is currently a frontier research area~\cite{AJLeggett_2002, MassiveQMreview2025, Knee2016}. In particular, direct tests of the quantum superposition  principle for the states of massive systems have so far been achived by quantum interference experiments with macromolecules of masses up to $\sim 10^{-22}$ kg~\cite{Fein2019, Pedalino2026}, and by preparing Schr\"odinger's cat state of a mechanical resonator of mass $\sim 10^{-9}$ kg mass~\cite{Bild2023}. These experiments directly test the quantum superposition principle in the macroscopic regime. Extending such experiments to even larger masses, however, is extremely challenging. In such studies, a major difficulty is that quantum effects in the macroscopic domain become very small due to unavoidable environmental interactions. Consequently, in order to display the macroscopic quantum mechanical effects, the conventional approaches primarily focus on testing macroscopic quantum superposition by suppressing environmental damping effects. Here, we pursue a complementary route by focusing on testing the other cardinal principle of quantum mechanics, namely, the inevitable disturbance induced by quantum measurements, whose experimental demonstration in the macroscopic domain remains comparatively unexplored. In this direction, an experiment~\cite{DasMassindependent2024} has recently been proposed based on testing the classical notion of macrorealism (MR)~\cite{AJLeggett_2002}. However, this proposal, too, is predicted to yield a tiny signature of nonclassicality even under ideal conditions. The key idea underpinning our present work is to amplify such measurement-induced quantumness for massive systems through a suitable sequence of repetitive measurements that can progressively inhibit the system's dynamical evolution.  This phenomenon is called the Quantum Zeno Effect (QZE)~\cite{Home1986, Home1992, 
Home1997, Facchi2004, Georgescu2022},  first formulated by Misra and Sudarshan~\cite{Misra1977}. In the present work, we propose a suitably controlled optomechanical experiment to witness the QZE-induced enhancement of nonclassicality for massive mechanical oscillators.

The first experimental demonstration of QZE was provided by Itano \textit{et al.}, showing the inhibition of atomic transitions through repeated measurements~\cite{Itano1990}. Since then, QZE has been extensively investigated, with comprehensive analyses and generalizations being presented in \cite{Home1992,Home1997, Facchi2004, Georgescu2022}. Beyond its foundational implications, QZE has emerged as a powerful tool in decoherence control~\cite{Facchi2002}, interaction-free measurements~\cite{Kwiat1995}, and the manipulation of dynamics in many-body systems~\cite{Signoles2014,BurgarthExponential2014}. Moreover, QZE has been explored for macroscopic quantum systems, such as Josephson junctions~\cite{Barone2004}, Bose--Einstein condensates~\cite{Shchesnovich2010,Barontini2013}, macroscopic two-level systems~\cite{Ghasemi2019}. In all these studies, macroscopicity is characterized by collective quantum degrees of freedom, while the physical mass of the quantum system itself is not used as the relevant macroscopic parameter. In contrast, we instead consider a massive quantum harmonic oscillator, treating its physical mass as the macroscopic parameter, which is the key parameter for  a number of important foundational applications, such as those related to the quantum nature of gravity \cite{BoseSpinwitness2017,Marletto2017,HanifTesting2024,MassiveQMreview2025,bosespin2025,Das2026How}, and the tesing of wave function collapse model~\cite{Ghirardi1986,Bassi2013, Donadi2021}. 

In this paper, we propose a near-term realisable scheme to harness the QZE in massive systems. We consider a coherent state evolving in a harmonic potential and subject it to repeated measurements that probe whether the system remains in its initial state. Coherent states are particularly suitable for this purpose, as they are minimum-uncertainty states that preserve their shape and uncertainty product during evolution, representing the most classical among quantum states while remaining experimentally realisable even for massive systems~\cite{Itano1997,Aspelmeyer2014}. The effect of successive measurements is quantified through the survival probability (i.e. the probability of finding the system in its initial state), a standard indicator of QZE~\cite{Misra1977,Home1992,Home1997}. By performing this sequence of measurements, we can directly examine how repeated observations influence the system’s evolution and exploit the cumulative effect of these disturbances to amplify signatures of quantum behavior that are otherwise suppressed in macroscopic systems. The scheme relies on trapping~\cite{Ballestero2021} and cooling~\cite{Gieseler2012} a mesoscopic mechanical object to its ground state, followed by a sequence of measurements. Each measurement consists of a beam-splitter-type light–matter interaction and a subsequent binary-outcome measurement on the optical mode corresponding to photon detection or no detection.

As regards quantification of the quantumness entailed by QZE, we first note a fundamental distinction between measurements in quantum physics and classical physics. In particular, ideal measurements on a classical system can, in principle, be fully noninvasive, leaving both the system’s evolution and all subsequent outcomes unaffected \cite{LeggettGarg1985,AJLeggett_2002}. Quantum measurements, in contrast, inevitably disturb the system, and the resulting disturbance cannot be eliminated, even under ideal conditions. The in-principle non-invasive nature of measurements in classical physics leads to the experimentally testable  \textit{non-disturbance condition} (NDC)~\cite{
Kofler2013,Schild2015,Knee2016}, whose loophole-free violation can demonstrate the genuine quantumness resulting from the quantum measurement-induced disturbance (wave function collapse) \cite{zindorfHow2025,bracciniNonclassicality2025}. Since QZE is a manifestation of measurement-induced quantum disturbance, quantum violation of the NDC involving a large number of sequential measurements provides a direct witness of nonclassicality through the cumulative effect of irreducible quantum disturbances induced by repeated quantum measurements, thereby enabling one to quantify the QZE. 
By connecting the survival probability of finding the system in the initial state under repeated measurements to the quantum violation of NDC, we quantitatively show that the QZE can be systematically harnessed to amplify the otherwise suppressed quantum effects caused by measurement-induced quantum disturbance, thus offering a practical and scalable route to detect macroscopic quantumness.

 A key feature of our formulated scheme is that appropriate control experiments have also been devised to determine quantitatively the effect of unwanted classical disturbance arising from implementing our proposed measurement protocol; importantly, this would enable an unambiguous operational identification of the inherent quantum component of the measurement-induced disturbance.


{\em Schematics.---} We consider a single quantum harmonic oscillator of angular frequency $\Omega_m$, prepared at time $t=0$ in the mechanical coherent state $\rho_{0}=|\alpha\rangle_m\langle\alpha|.$
To probe the influence of repeated measurements on its evolution, we divide the total evolution time into intervals of duration $\delta t$. At the end of each interval, the system undergoes 
the measurement
$\{E_{0}=|\alpha\rangle_m\langle\alpha|, E_{1}=\mathbb{I}_m-|\alpha\rangle_m\langle\alpha|\}$
with outcomes $0$ and $1$ assigned to $E_{0}$ and $E_{1}$ respectively. The first measurement takes place at $t=\delta t$, after which the system evolves freely under the harmonic potential until the next step. This evolution–measurement cycle is repeated $N-1$ times, followed by a final measurement at $t=N\delta t=t_f$. This enables a periodic interrogation of the coherent-state dynamics. 

In this framework, the \textit{survival probability} quantifies the likelihood that the final measurement at time $t_f$ yields the outcome $0$, independent of the specific sequence of outcomes obtained at earlier stages. All intermediate results are therefore summed over. Let us denote the outcomes of the measurements at times $\delta t,2\delta t,\ldots,(N-1)\delta t$ by $a_1,\ldots,a_{N-1}\in\{0,1\}$,  and the joint probability of obtaining the sequence of outcomes
$(a_1,\ldots,a_{N-1},a_N=0)$ by
$P(a_1,\ldots,a_{N-1},a_N=0)$. The \textit{survival probability} in the presence of intermediate measurements is given by
\begin{equation}
P_{0}^{\text{(with)}}(t_f)
= \sum_{a_1,\ldots,a_{N-1}=0,1}
P(a_1,a_2,\ldots,a_{N-1},a_N = 0).
\end{equation}
This quantity captures the cumulative effect of repeated measurements on the system's evolution and serves as the operational figure of merit in the Quantum Zeno protocol.

In another version of the setup, the system is allowed to evolve uninterrupted after its preparation at $t=0$. No measurement is introduced during the interval $(0,t_f)$, so the dynamics proceed solely under the harmonic potential. At the final time $t=t_f$, a single measurement, identical to the one used in the monitored setting, is performed. The probability that this measurement yields the outcome $0$ defines the corresponding survival probability, which we denote by 
$P_{0}^{\text{(without)}}(t_f)$.

The \textit{no-disturbance condition} serves as a classical benchmark, implying that the intermediate measurements, whose outcomes are later discarded, should leave the final statistics unchanged. For the scheme considered here, this condition takes the form,
\begin{equation}
\kappa_{\mathrm{NDC}}
= 0, 
 \label{KNDC}
\end{equation}
where
\begin{equation}
\kappa_{\mathrm{NDC}}
= P_{0}^{\text{(with)}}(t_f)
 - P_{0}^{\text{(without)}}(t_f).   
 \label{KNDCform}
\end{equation}

Any deviation from this relation, $\kappa_{\mathrm{NDC}}\neq 0$, signals that the intermediate measurements have disturbed the system’s evolution. The quantity $\kappa_{\mathrm{NDC}}$ thus offers a direct, operational measure of measurement-induced disturbance and serves as an indicator of nonclassical dynamics in the monitored oscillator.

Note that if, for a large number of sequential measurements, $P_{0}^{\text{(with)}}(t_f)\rightarrow 1$ and $P_{0}^{\text{(without)}}(t_f) \rightarrow 0$, then we have $\kappa_{\mathrm{NDC}} \rightarrow 1$, indicating effective freezing of the dynamics in this limiting condition (the original formulation of the QZE). This entails an increasing magnitude of NDC violation induced by the QZE as the number of measurements increases.
  

\begin{figure*}[t]
\centering

\begin{subfigure}[t]{0.48\textwidth}
    \centering
    \includegraphics[width=\linewidth]{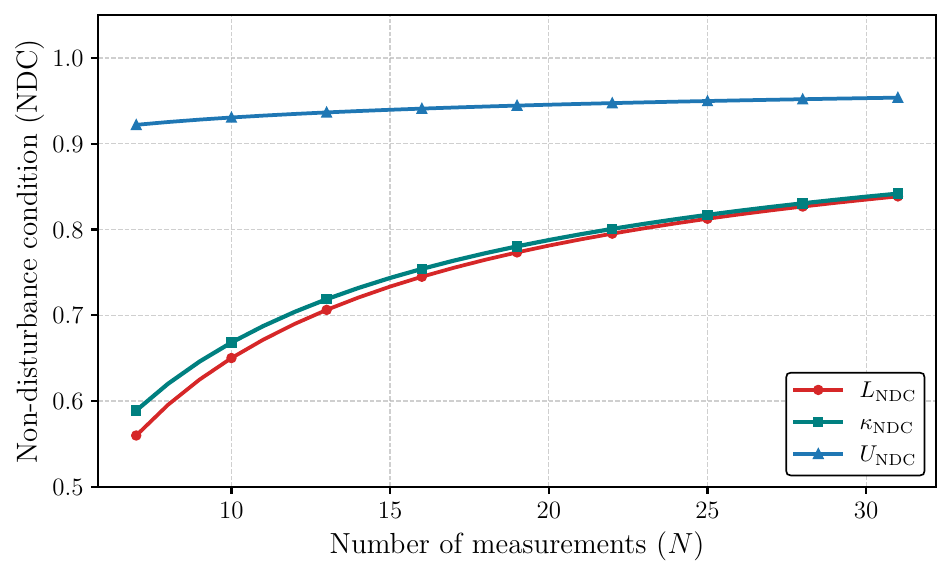}
    \caption{}
\end{subfigure}
\hfill
\begin{subfigure}[t]{0.48\textwidth}
    \centering
    \includegraphics[width=\linewidth]{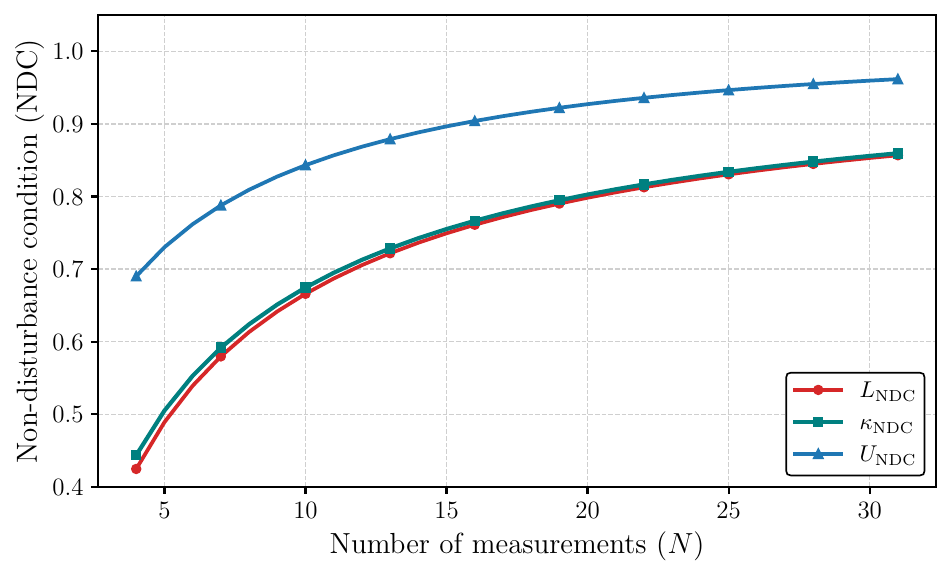}
    \caption{}
\end{subfigure}

\caption{\footnotesize
The blue curve shows the upper limit
$U_{\mathrm{NDC}}$, the red curve shows the lower limit
$L_{\mathrm{NDC}}$, and the green curve shows the numerically
computed value of $\kappa_{\mathrm{NDC}}$. Panels (a) and
(b) correspond to:
(a) $\Omega_m t_f = \frac{\pi}{2}$ and
$|\alpha| = N^{1/16}$ for $7 \le N \le 100$,
(b) $\Omega_m t_f = \frac{\pi}{4}$ and
$|\alpha| = N^{1/4}$ for $4 \le N \le 100$.
In both cases, $(N-1)$ is the number of intermediate
measurements, and the actual NDC violation lies between
the upper and lower bounds.
}
\label{NDC-combined}
\end{figure*}

{\em Analysis.}--- We begin our analysis by considering that the mechanical oscillator evolves under the harmonic Hamiltonian
$\hat{H} = \hbar \Omega_m (\hat{n} + \mathbb{I}/2)$ for a total duration $t_f$,
followed by the final measurement $\{E_0,E_1\}$. Under harmonic evolution, the state at time $t_f$ is
\begin{equation}
\rho_{t_f} = |\alpha e^{-i\Omega_m t_f}\rangle_m\langle \alpha e^{-i\Omega_m t_f}|.
\end{equation}
In the absence of intermediate measurements, the probability of obtaining the
outcome $0$ in the final measurement is
\begin{align}
P_{0}^{\text{(without)}}(t_f)
&= \mathrm{Tr}(\rho_{t_f} E_0) \nonumber \\
&= \exp\left[-2|\alpha|^2\bigl(1-\cos\Omega_m t_f\bigr)\right].
\label{Pwithout}
\end{align}
Throughout our analysis, we fix the total evolution phase as
$\Omega_m t_f = \Omega_m N \delta t \equiv \theta$, where $\theta>0$ is a finite
constant independent of $N$. This requires that, in the large-$N$ limit, the interval $\delta t$ is reduced such that the product $\Omega_m N \delta t = \theta$ remains fixed for all $N$.

\begin{figure*}[t]
\centering
\includegraphics[width=0.8\linewidth]{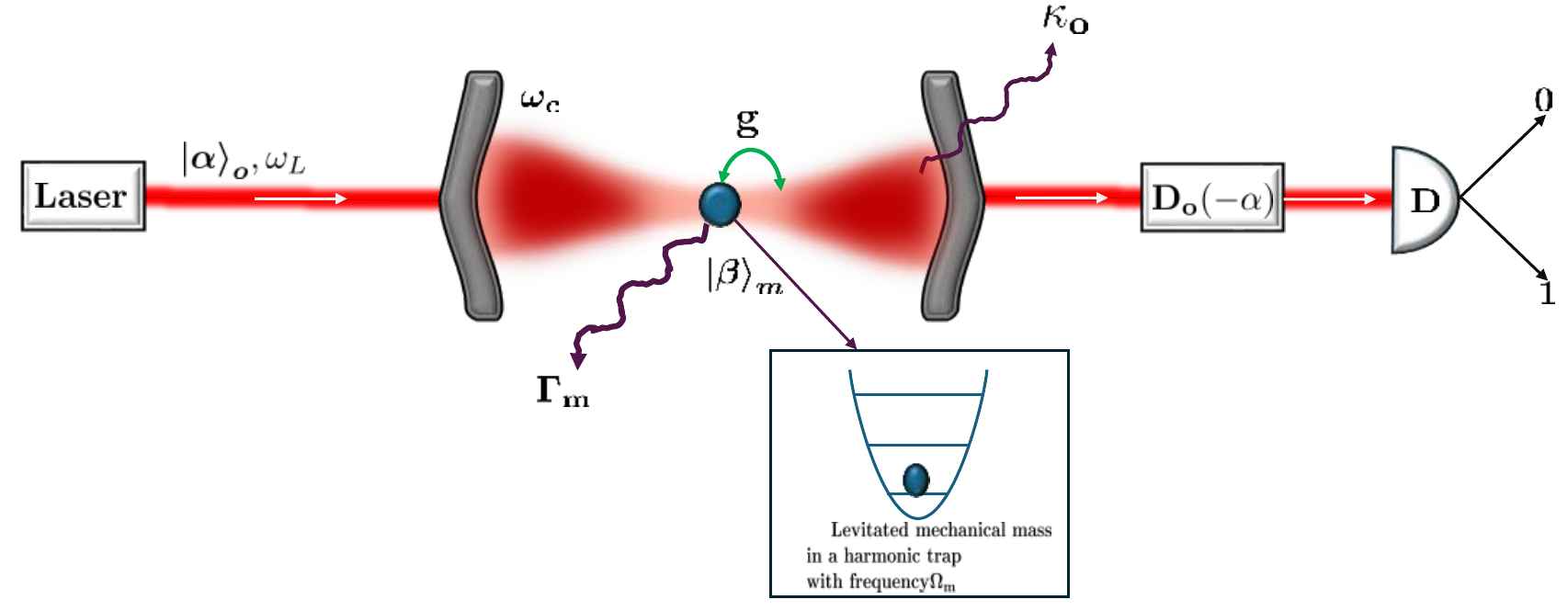}

\caption{\footnotesize
Levitated and suitably trapped mechanical mass $m$, trapped in a harmonic potential with mechanical frequency $\Omega_m$, is placed inside a cavity of frequency $\omega_c$, driven by input laser of state $\ket{\alpha}_o$ and frequency $\omega_L$. The output from cavity is subjected to an optical displacement $D_o (-\alpha)$, followed by a on/off photon detector (D). Here, $g$ denotes the optomechanical coupling, $\kappa_o$ is the optical cavity decay rate, and $\Gamma_m$ denotes the mechanical damping rate. The cavity is red-detuned by setting $\omega_L - \omega_c = - \Omega_m$, the resulting linearized optomechanical interaction realizes a beam-splitter Hamiltonian.}
\label{fig:opto_expt}

\end{figure*}

We next consider the scenario in which repeated measurements
$\{E_0,E_1\}$ are performed between successive harmonic evolutions of
duration $\delta t$, starting from the same initial state $\rho_0$.
To evaluate $\kappa_{\text{NDC}}$ in Eq.~(\ref{KNDC}), we compute the
probability of obtaining the outcome $0$ at the final measurement,
\begin{align}
P_{0}^{\text{(with)}}(t_f)
&= \sum_{a_1,\ldots,a_{N-1}=0,1}
P(a_1,a_2,\ldots,a_{N-1},a_N=0) \nonumber\\
&\ge P(a_1=0,a_2=0,\ldots,a_{N-1}=0,a_N=0),
\label{PwithT}
\end{align}
where the inequality follows by retaining only the contribution corresponding
to the outcome $0$ in all $N$ measurements.
The quantity
$P(a_1=0,a_2=0,\ldots,a_{N-1}=0,a_N=0)$ denotes the joint probability of
obtaining an outcome $0$ at every measurement step.
In the following, we evaluate this joint probability explicitly and use it to
derive a lower bound on $P_{0}^{\text{(with)}}(t_f)$, which provides a testable constraint on $\kappa_{\text{NDC}}$.

To evaluate the lower bound of joint probability analytically in the large $N$ regime, we consider the limit of short-time evolution between successive measurements. Assuming $\Omega_m \delta t \ll 1$, we expand the harmonic evolution to second order as, $\cos(\Omega_m \delta t) \simeq 1 - \frac{\Omega_m^2(\delta t)^2}{2}$. Within this approximation (see Supplemental Material~\cite{Lzeno_SM} for details), the joint probability of obtaining the outcome $0$ in all $N$ measurements becomes
\begin{align}
P(a_1=0,\ldots,a_N=0)
&= \left[\exp\left(-2|\alpha|^2\bigl(1-\cos\Omega_m\delta t\bigr)\right)\right]^N \nonumber\\
&\simeq \exp\left[-\frac{|\alpha|^2\theta^2}{N}\right],
\label{jointprob}
\end{align}
where we have used $t_f =N\delta t$ and $\Omega_m t_f=\theta$. As a result, we get
\begin{equation}\label{eq:P_with_approx}
P_{0}^{\text{(with)}}(t_f)
\ge \exp\left[-\frac{|\alpha|^2\theta^2}{N}\right],
\end{equation}
which leads to the bound
\begin{equation}
\kappa_{\mathrm{NDC}}
\ge \exp\left[-\frac{|\alpha|^2\theta^2}{N}\right]
- \exp\left[-2|\alpha|^2(1-\cos\theta)\right].
\label{ndcbound}
\end{equation}

We focus on finite values of $\theta$ for which $1-\cos\theta$ remains finite and assume a sublinear scaling of the coherent-state amplitude with the number of measurements,
$|\alpha| = c N^{1/x}$ with $c>0$ and $x>2$. Such choices of $\theta$ and $|\alpha|$ ensures that $\kappa_{\mathrm{NDC}} \to 1$ when $N \to \infty$ (which is equivalent to QZE) as shown below.
With these choices, the lower bound in Eq.~(\ref{ndcbound}) implies
\begin{align}
\lim_{N\to\infty} P_{0}^{\text{(with)}}(t_f) = 1,
\label{afterintN}
\end{align}
where we have used the fact that a valid probability cannot be greater than unity. On the other hand, in the absence of intermediate measurements vanishes,
\begin{align}
\lim_{N\to\infty} P_{0}^{\text{(without)}}(t_f)
&= \lim_{N\to\infty} \exp\!\left[-2|\alpha|^2(1-\cos\theta)\right] \nonumber\\
&= \lim_{N\to\infty} \exp\!\left[-2c^2 N^{2/x}(1-\cos\theta)\right] \nonumber\\
&= 0.
\label{withoutintN}
\end{align}
Combining Eqs.~(\ref{afterintN}) and (\ref{withoutintN}), we obtain
\begin{equation}
\lim_{N\to\infty} \kappa_{\mathrm{NDC}} = 1.
\label{limndc}
\end{equation}
Thus, infinitely frequent measurements inhibit the harmonic evolution, confining the system to its initial state at $t=t_f$, while the survival probability vanishes without intermediate measurements. This behavior is a manifestation of the QZE.

\begin{figure*}[t]
\centering

\begin{subfigure}[t]{0.49\textwidth}
    \centering
    \includegraphics[width=\linewidth]{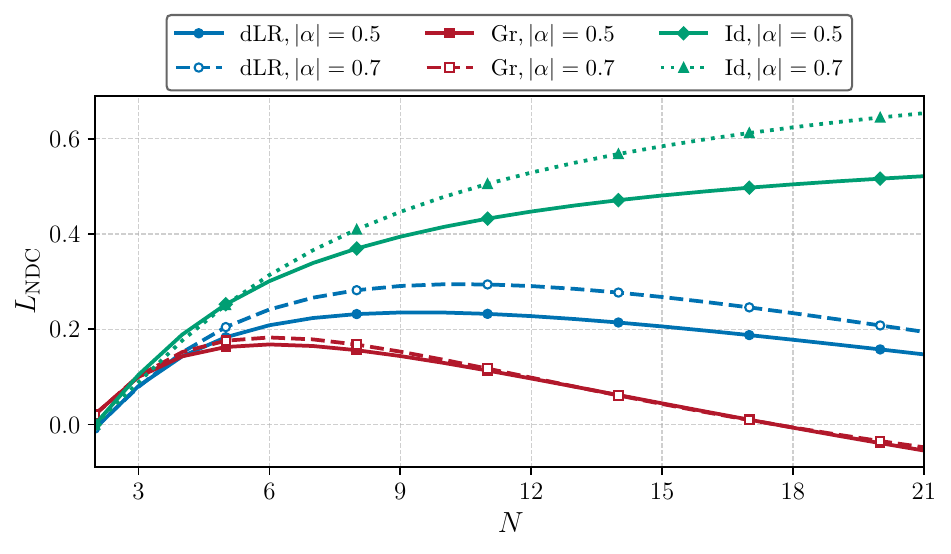}
    \caption{}
    \label{fig:real_ndc}
\end{subfigure}
\hfill
\begin{subfigure}[t]{0.49\textwidth}
    \centering
    \includegraphics[width=\linewidth]{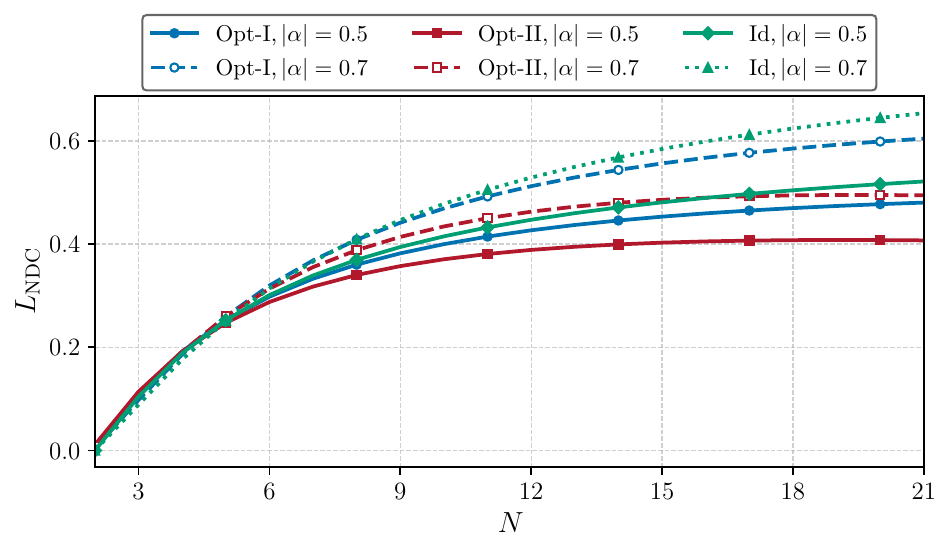}
    \caption{}
    \label{fig:optimistic_ndc}
\end{subfigure}

\caption{\footnotesize
Non-Disturbance Condition (NDC) lower bound $L_{\rm NDC}$ enhancement under sequential operations
as a function of the number of operations $N$.
(a) For experimentally realized optomechanical parameter regimes with $m \sim 10^{-18}$ kg and $m \sim 10^{-10}$ kg:
de los Ríos \textit{et al.} (dLR) ~\cite{deLosRiosSommer2021} and Gr\"oblacher \textit{et al.} (Gr)~\cite{Groblacher2009}.
(b) Performance in optimistic parameter regimes (Opt-I and Opt-II)
with a moderate improvement over the experimentally demonstrated optomechanical coupling ($g$)-to-optical cavity decay rate ($\kappa_o$) ratio, while keeping all other parameters in the experimentally realizable regime \cite{deLosRiosSommer2021,Groblacher2009}.
Solid and dashed curves correspond to coherent amplitudes
$|\alpha|=0.5$ and $|\alpha|=0.7$, respectively.
The ideal lossless evolution (Id) is shown as a reference.
}
\label{fig:damped_ideal_ndc}

\end{figure*}
In any realistic experimental implementation, the number of intermediate measurements is necessarily finite, making it essential to quantify the attainable NDC violation away from the ideal Zeno limit. Although the exact value of $\kappa_{\mathrm{NDC}}$ cannot be obtained analytically for finite $N$, it can be rigorously constrained within analytically tractable bounds. We therefore introduce lower and upper bounds, $L_{\mathrm{NDC}}$ and $U_{\mathrm{NDC}}$, such that
$U_{\mathrm{NDC}} \ge \kappa_{\mathrm{NDC}} \ge L_{\mathrm{NDC}}$.

From Eq.~(\ref{ndcbound}), the lower bound is given by
\begin{equation}
L_{\mathrm{NDC}}
= \exp\left[-\frac{|\alpha|^2\theta^2}{N}\right]
- \exp\left[-2|\alpha|^2(1-\cos\theta)\right],
\label{Klower}
\end{equation}
while, from Eqs.(\ref{KNDCform}) and (\ref{Pwithout}), the upper bound is given by,
\begin{equation}
U_{\mathrm{NDC}}
= 1 - \exp\!\left[-2|\alpha|^2(1-\cos\theta)\right].
\label{Kupper}
\end{equation}

These bounds constrain the attainable NDC violation for finite $N$; however, determining 
$\kappa_{\mathrm{NDC}}$ quantitatively requires resolving the measurement-conditioned dynamics. We emphasize the distinction between the joint probability
$P(a_1=0, a_2=0,\ldots,a_N=0)$, associated with a specific measurement record, and the survival probability
$P_{0}^{\text{(with)}}(t_f)$, which gives the probability of obtaining an outcome $0$ at the final measurement,
independent of the sequence of intermediate outcomes and hence summed over all measurement histories.
This distinction is immaterial in the Zeno limit but becomes essential at finite $N$.
Since a closed-form analytical expression for $P_{0}^{\text{(with)}}(t_f)$ is highly nontrivial,
we evaluate $\kappa_{\mathrm{NDC}}$ numerically using a truncated coherent-state representation of the oscillator. Details of this numerical approach is provided in the  Supplemental Material~\cite{Lzeno_SM}.
This approach faithfully captures the relevant Fock-space support of the initial state and yields accurate
finite-$N$ estimates of $\kappa_{\mathrm{NDC}}$, which are found to lie within the bounds
$L_{\mathrm{NDC}}$ and $U_{\mathrm{NDC}}$. See Fig.\ref{NDC-combined} for details.

\textit{Possible experimental implementation.---} The protocol can be implemented using a harmonically trapped mechanical oscillator, as routinely realised in levitated optomechanical platforms~\cite{Ballestero2021}. Possible implementations include mesoscopic mechanical oscillators based on levitated optomechanics in low-noise optical dipole, ion, magnetic, or diamagnetic traps operated under ultrahigh vacuum and cryogenic conditions. 
The mechanical mode is first cooled close to its motional ground state using feedback-cooling techniques~\cite{Gieseler2012}, where the cooling rate exceeds the environmental heating rate~\cite{Doherty1999, Walker2019, Vinante2019}, and followed by a coherent displacement to prepare our desired initial coherent state.

We propose a cavity-optomechanical realization of the
two-outcome measurement described by the POVM
$\{E_0=|\alpha\rangle_m\langle\alpha|,
E_1=\mathbb{I}_m-|\alpha\rangle_m\langle\alpha|\}$.
Let $a$ ($a^\dagger$) and $b$ ($b^\dagger$) denote the optical
and mechanical annihilation (creation) operators, respectively.
Under red detuning and in the resolved-sideband regime $\Omega_m\gg\kappa_o$ (where $g$ is the optomechanical coupling and $\kappa_o$ is the optical cavity decay rate), and within the rotating-wave approximation ($g\ll\Omega_m$), the interaction reduces to the beam-splitter Hamiltonian,
$H=\hbar\Omega_m(a^\dagger a+b^\dagger b)
-\hbar g(a^\dagger b e^{-i\pi/2}
+ab^\dagger e^{i\pi/2})$~\cite{Aspelmeyer2014}.

Fig.~\ref{fig:opto_expt} illustrates how our proposed setup works. Let the mechnaical oscillator be in an arbitrary coherent state $|\beta\rangle_m$. The optical ancilla is initialized in the coherent state
$|\alpha\rangle_o$ and interacts with the mechanical mode for
$t_{\rm swap}=\pi/(2g)$. In the strong-coupling regime
$g\gg\kappa_o,\Gamma_m$ ($\Gamma_m$ being the mechanical damping rate), this realizes a coherent state swap.
In the ideal dissipation-free limit one obtains an ideal swap:
$|\alpha\rangle_o|\beta\rangle_m
\rightarrow|\beta\rangle_o|\alpha\rangle_m$.
Including weak dissipation, the swap becomes
$|\alpha\rangle_o|\beta\rangle_m
\rightarrow |A\beta\rangle_o|A\alpha\rangle_m$,
where
$A=\exp[-(\kappa_o+\Gamma_m)t_{\rm swap}/4]
e^{-i\Omega_m t_{\rm swap}}$
(see~\cite{Lzeno_SM}).

After the swap, an optical displacement
$D_o(-\alpha)$ is applied, and the optical mode is measured using
an on/off photo-detector defined by the projective measurement
$\{|0\rangle_o\langle0|,
\mathbb{I}_o-|0\rangle_o\langle0|\}$.
In the ideal limit the no-click probability is
$P_{\rm no}=|\langle\alpha|\beta\rangle|^2$,
whereas weak dissipation modifies it to
$P_{\rm no}=\exp[-|A\beta-\alpha|^2]$.

Conditioned on the no-click outcome, the mechanical update is
described by the Kraus operator
$K_{\rm no}=|\alpha\rangle_m\langle\alpha|$, giving
$E_0=K_{\rm no}^{\dagger}K_{\rm no}
=|\alpha\rangle_m\langle\alpha|$.
Together with the complementary click branch satisfying
$E_1=K_{\rm click}^{\dagger}K_{\rm click}
=\mathbb{I}_m-|\alpha\rangle_m\langle\alpha|$,
this realizes the required two-outcome measurement
instrument. Note that the analytically derived upper and lower bounds of the $\kappa_{\text{NDC}}$ are valid for this implementation of the measurements, independent of any realization of $K_{\rm click}$. The damping analysis is also performed for this implementation\footnote{Note that the numerical evaluation of
$\kappa_{\mathrm{NDC}}$ is performed for a specific $\{K_{\rm no}, K_{\rm click}\}$. The possible implementation of this measurement instrument is slightly different from the above protocol, and is presented in~\cite{Lzeno_SM}.}.

We now show that the protocol remains valid for an arbitrary mechanical state. Any pure state of the mechanical oscillator admits a coherent-state expansion
$|\psi\rangle_m=\frac{1}{\pi}\int d^2\beta\,\langle\beta|\psi\rangle\,|\beta\rangle_m$.
Since the protocol is independent of the input coherent state, $K_{\rm no}$ acts linearly on the above expansion, giving
\begin{align}
K_{\rm no}|\psi\rangle_m
&=\frac{1}{\pi}\int d^2\beta\,\langle\beta|\psi\rangle\,
K_{\rm no}|\beta\rangle_m =|\alpha\rangle_m\langle\alpha|\psi\rangle.
\end{align}
Thus, $K_{\rm no}=|\alpha\rangle\langle\alpha|$ for an arbitrary pure state of the mechanical oscillator. Since quantum operations act linearly on density operators, the protocol also extends immediately to arbitrary mixed states. This proves that the aforementioned protocol realises the measurement $\{E_{0}=|\alpha\rangle_m\langle\alpha|, E_{1}=\mathbb{I}_m-|\alpha\rangle_m\langle\alpha|\}$ on the mechanical oscillator at any stage, even when the mechanical oscillator is not in a pure coherent state.

{\it Damping analysis.}--- For the realistic dissipative analysis, we evaluate the experimentally relevant lower bound on $\kappa_{\mathrm{NDC}}$ ($L_{\mathrm{NDC}}$), obtained from the trajectory in which every measurement yields the no-click outcome. Since $L_{\mathrm{NDC}}$ is sufficient to certify measurement-induced non-classicality even in the presence of damping, we restrict our analysis to this quantity, which also considerably simplifies the theoretical analysis. Hence, only the $K_{\rm no}$ branch enters the
damped evolution, as follows from Eq.~\eqref{Klower}. Each measurement cycle consists of free
mechanical evolution for $\delta t_{\rm free}$ followed by one
optomechanical swap and optical readout. Therefore,
$\delta t=\delta t_{\rm free}+t_{\rm swap}$ and
$t_f=N\delta t$, with coherent operation requiring
$t_f\ll\Gamma_m^{-1}$. The total evolution time is chosen to maximize this lower bound by minimizing the single-shot no-click probability while keeping the sequential no-click probability high. We therefore fix $\Omega_m t_f=(2k+1)\pi$, where the integer $k$ satisfies $\frac{1}{2}\!\left(\frac{\Omega_m N}{2g}-1\right)\lesssim k\lesssim\frac{1}{2}\!\left(\frac{\Omega_m}{\pi\Gamma_m}-1\right)$, following from the constraints $N t_{\rm swap}<t_f\ll\Gamma_m^{-1}$. Furthermore, imposing the phase-matching condition $\Omega_m t_{\rm swap}=2\pi q$ ($q\in\mathbb{N}$), we choose the smallest compatible value, namely $k=qN$. This ensures that the measurement interval $t_f/N$ lies in the quantum Zeno regime for sufficiently large $N$~(see~\cite{Lzeno_SM}).

For the damping analysis, we consider both experimentally
motivated and optimistic optomechanical regimes. The mechanical
frequency is chosen to satisfy the resonance condition
$\Omega_m=4g$ for all cases. The experimentally motivated
parameters are inspired by
de~los~R\'{\i}os~Sommer
\textit{et al.}~\cite{deLosRiosSommer2021}, with
$m=6.4\times10^{-18}\,\mathrm{kg}$,
$\Omega_m/g=4$, $\Omega_m/\Gamma_m=100$, and
$g/\kappa_o=2.28$. We also consider the parameters of
Gr\"oblacher \textit{et al.}~\cite{Groblacher2009},
with $m=1.45\times10^{-10}\,\mathrm{kg}$,
$\Omega_m/g=4$,
$\Omega_m/\Gamma_m\simeq6.76\times10^3$, and
$g/\kappa_o\simeq1.51$.

To explore improved future devices, we additionally consider two
optimistic regimes. Optimistic-I assumes
$\Omega_m/g=4$, $\Omega_m/\Gamma_m=3.0\times10^3$, and
$g/\kappa_o=10$, representing reduced optical loss. Optimistic-II takes
$\Omega_m/g=4$, $\Omega_m/\Gamma_m=3.0\times10^3$, and
ratio $g/\kappa_o=5$. The resulting lower bound of the dissipative
$\kappa_{\mathrm{NDC}}$ is shown in
Fig.~\ref{fig:damped_ideal_ndc}. A comparison between the experimentally motivated Fig.~\ref{fig:damped_ideal_ndc}(a) and optimistic Fig.~\ref{fig:damped_ideal_ndc}(b) regimes shows that just moderate improvements in the optomechanical coupling-to-optical cavity decay rate ratio $g/\kappa_o$ lead to a significant enhancement of $\kappa_{\mathrm{NDC}}$, bringing the dissipative dynamics much closer to the ideal lossless limits.  

Note that Fig.~\ref{fig:damped_ideal_ndc} presents finite-$N$ results for fixed coherent-state amplitudes ($|\alpha|=0.5$ and $0.7$). While we assumed earlier (also in Fig. \ref{NDC-combined}) that $|\alpha| = c N^{1/x}$ with $c>0$ and $x>2$ to analytically demonstrate perfect QZE, i.e., $\kappa_{\mathrm{NDC}} \to 1$ when $N \to \infty$, we have considered smaller fixed values of $|\alpha|$ here to simplify the numerical calculations. Although this choice does not exhibit perfect QZE, it still demonstrates a significant enhancement of $\kappa_{\mathrm{NDC}}$ with increasing $N$ (to some extent), even in the presence of realistic damping.

{\em Classical disturbance calibration.---}
The experimentally measured witness, $\kappa_{\mathrm{NDC}}^{\mathrm{exp}}$, generally contains both the desired quantum contribution arising from quantum measurement-induced disturbance and residual classical disturbance (CD) arising from the effect solely due to the presence of the optical probe, detector imperfections, and other technical noise. To estimate this classical contribution, we propose two control experiments using the same measurement protocol as in the main experiment. First, the mechanical oscillator is initialized in the vacuum state $|0\rangle_m$. The associated measurement,
$\{|0\rangle_m\langle0|,\mathbb{I}_m-|0\rangle_m\langle0|\}$,
is ideally nondisturbing for this state. Since the vacuum is also stationary under harmonic evolution, quantum mechanics predicts identical final probabilities with and without the intermediate measurements. Second, the oscillator is prepared in a coherent state $|\alpha\rangle_m$ and allowed to evolve for integer multiples of its oscillation time period before each measurement. As the coherent state returns to its initial state $|\alpha\rangle_m$ after every complete period, the measurement $\{|\alpha\rangle_m\langle\alpha|,\mathbb{I}_m-|\alpha\rangle_m\langle\alpha|\}$ always yields the no-click outcome in the ideal case. Thus, both control experiments ideally give $\kappa_{\mathrm{NDC}}=0$, and any observed nonzero value is attributed to classical disturbance, which we denote by $\kappa_{\mathrm{NDC}}^{\mathrm{CD}}$. The genuine quantum contribution is then obtained as
\begin{equation}
\kappa_{\mathrm{NDC}}^{\mathrm{QM}}
=
\kappa_{\mathrm{NDC}}^{\mathrm{exp}}
-
\kappa_{\mathrm{NDC}}^{\mathrm{CD}}.
\end{equation}
A nonzero $\kappa_{\mathrm{NDC}}^{\mathrm{QM}}$ therefore certifies a genuine quantumness after eliminating the contribution from classical disturbances.

{\em Conclusion.---}  In the pursuit of testing quantumness of macroscopic objects, the studies to date have largely sought to probe various aspects of the quantum superposition principle in the macroscopic domain by focusing on mitigating the decohering effects which tend to mask the usually very small quantitative signatures of macroquantumness. In contrast, our present work seeks to shift the focus towards formulating a suitable strategy for enhancing quantitatively the macroquantum effect in itself, thereby facilitating its detection even in the presence of strong decohering effects. For this purpose, we propose a realizable experimental scheme that, for the first time, invokes the QZE for massive systems, thereby enabling an appreciable enhancement of the observable signature of measurement-induced quantum disturbance, quantified by the violation of NDC. Here the key idea is that, although each individual measurement of finding whether the system is in the initial state or not, may induce only a tiny amount of quantum disturbance, the cumulative effects of many such measurements can lead to an appreciably enhanced manifestation of the resulting  quantum disturbance, the amount of such amplification being controllable by the chosen number of successive measurements.  However, a probable loophole in this scheme arises from the possibility of unavoidable classical disturbances occurring in the actual implementation of the proposed sequence of measurements. One of the highlights of our present paper is that to close this loophole, we have designed control experiments to determine the contribution from the associated classical disturbances to the measured violation of NDC. This would enable an unambiguous identification of the genuine quantum contribution to the observed violation of NDC. Our scheme requires trapping and ground-state cooling of the mass, without requiring the preparation and preservation of fragile macroscopic superposition states. The successive measurements can be implemented via beam-splitter-type interactions with an optical field, followed by photon detection. This scheme is realisable for mechanical oscillators of masses in the range $m \sim 10^{-10}-10^{-18}$ kg with near-term technologies, where strong signatures of Zeno-induced observable quantumness are expected even with realistic decoherence parameters. 

Several interesting directions remain open for future investigation. Optimisation of the measurement protocol may enable a more robust macroscopic quantum Zeno effect. More broadly, extending such a sequential measurement-based approach to other fundamental tests of physics, including  collapse models~\cite{Ghirardi1986,Bassi2013, Donadi2021}, gravity-induced decoherence scenarios~\cite{Diosi1987,Penrose1996}, and the quantumness of gravity~\cite{BoseSpinwitness2017,Marletto2017,HanifTesting2024} is worth pursuing for future research. Beyond its foundational significance, our protocol may also be useful for future sequential quantum metrology and sensing~\cite{Montenegro2022,Yang2023} in optomechanical systems.

\begin{acknowledgments}
{\it Acknowledgements.--}
DD and SB acknowledge the financial support from the Royal Society, UK under the scheme ``Newton International Fellowships
Alumni 2025'' (Grant No. AL$\backslash$251043). DD gratefully acknowledges University College London for its kind hospitality during his visit in July 2026, which provided the opportunity to complete a substantial part of this work. PR acknowledges the financial support from SNBNCBS, Kolkata. 
MT acknowledges funding from the Slovenian Research and Innovation
Agency (ARIS) under contracts N1-0392, P1-0416, SN-ZRD/22-27/0510
(RSUL Toro\v{s}).
HU acknowledges funding from the EU Horizon Europe EIC Pathfinder project QuCoM (10032223), from the UK funding agency EPSRC (grants  EP/V035975/1, EP/V000624/1, EP/W007444/1, EP/X009491/1), and from the Leverhulme Trust (RPG-2022-57). DH also acknowledges
support from NASI Senior Scientist Fellowship in initiating
this collaboration at Bose Institute, Kolkata. In completing
this work, DH acknowledges support as distinguished visitor
to the Light and Matter Physics group, RRI, Bangalore, under
the National Quantum Mission of the DST. SB would like to
acknowledge EPSRC grant EP/X009467/1 and STFC grant
ST/W006227/1. This work was made possible through
the support of the WOST, WithOut SpaceTime project
(https://withoutspacetime.org), supported by Grant ID\#63683
from the John Templeton Foundation (JTF). S.B’s research is
funded by the Gordon and Betty Moore Foundation through
Grant GBMF12328, DOI 10.37807/GBMF12328, and the Alfred P. Sloan Foundation under Grant No. G-2023-21130.
\end{acknowledgments}

\bibliography{ref}

@misc{Lzeno_SM,
  note = {See Supplemental Material for: (i) analytical derivations of the upper and lower bounds on $\kappa_{\mathrm{NDC}}$; (ii) numerical evaluation of $\kappa_{\mathrm{NDC}}$; (iii) an alternative implementation of the complete measurement instrument; and (iv) detailed calculations for a realistic experimental implementation under environmental damping.}
}

@article{AJLeggett_2002,
doi = {10.1088/0953-8984/14/15/201},
url = {https://doi.org/10.1088/0953-8984/14/15/201},
year = {2002},
month = {apr},
publisher = {},
volume = {14},
number = {15},
pages = {R415},
author = {A J Leggett},
title = {Testing the limits of quantum
mechanics:  motivation, state of play, prospects},
journal = {Journal of Physics: Condensed Matter}
}

@article{Ghirardi1986,
  title = {Unified dynamics for microscopic and macroscopic systems},
  author = {Ghirardi, G. C. and Rimini, A. and Weber, T.},
  journal = {Phys. Rev. D},
  volume = {34},
  issue = {2},
  pages = {470--491},
  numpages = {0},
  year = {1986},
  month = {Jul},
  publisher = {American Physical Society},
  doi = {10.1103/PhysRevD.34.470},
  url = {https://link.aps.org/doi/10.1103/PhysRevD.34.470}
}

@article{Bassi2013,
  title = {Models of wave-function collapse, underlying theories, and experimental tests},
  author = {Bassi, Angelo and Lochan, Kinjalk and Satin, Seema and Singh, Tejinder P. and Ulbricht, Hendrik},
  journal = {Rev. Mod. Phys.},
  volume = {85},
  issue = {2},
  pages = {471--527},
  numpages = {0},
  year = {2013},
  month = {Apr},
  publisher = {American Physical Society},
  doi = {10.1103/RevModPhys.85.471},
  url = {https://link.aps.org/doi/10.1103/RevModPhys.85.471}
}

@article{BoseSpinwitness2017,
  title = {Spin Entanglement Witness for Quantum Gravity},
  author = {Bose, Sougato and Mazumdar, Anupam and Morley, Gavin W. and Ulbricht, Hendrik and Toro\ifmmode \check{s}\else \v{s}\fi{}, Marko and Paternostro, Mauro and Geraci, Andrew A. and Barker, Peter F. and Kim, M. S. and Milburn, Gerard},
  journal = {Phys. Rev. Lett.},
  volume = {119},
  issue = {24},
  pages = {240401},
  numpages = {6},
  year = {2017},
  month = {Dec},
  publisher = {American Physical Society},
  doi = {10.1103/PhysRevLett.119.240401},
  url = {https://link.aps.org/doi/10.1103/PhysRevLett.119.240401}
}

@article{Marletto2017,
  title = {Gravitationally Induced Entanglement between Two Massive Particles is Sufficient Evidence of Quantum Effects in Gravity},
  author = {Marletto, C. and Vedral, V.},
  journal = {Phys. Rev. Lett.},
  volume = {119},
  issue = {24},
  pages = {240402},
  numpages = {5},
  year = {2017},
  month = {Dec},
  publisher = {American Physical Society},
  doi = {10.1103/PhysRevLett.119.240402},
  url = {https://link.aps.org/doi/10.1103/PhysRevLett.119.240402}
}

@article{LeggettGarg1985,
  title = {Quantum mechanics versus macroscopic realism: Is the flux there when nobody looks?},
  author = {Leggett, A. J. and Garg, Anupam},
  journal = {Phys. Rev. Lett.},
  volume = {54},
  issue = {9},
  pages = {857--860},
  numpages = {0},
  year = {1985},
  month = {Mar},
  publisher = {American Physical Society},
  doi = {10.1103/PhysRevLett.54.857},
  url = {https://link.aps.org/doi/10.1103/PhysRevLett.54.857}
}

@article{Misra1977,
    author = {Misra, B. and Sudarshan, E. C. G.},
    title = {The Zeno’s paradox in quantum theory},
    journal = {Journal of Mathematical Physics},
    volume = {18},
    number = {4},
    pages = {756-763},
    year = {1977},
    month = {04},
    issn = {0022-2488},
    doi = {10.1063/1.523304},
    url = {https://doi.org/10.1063/1.523304}
}

@article{Itano1990,
  title = {Quantum Zeno effect},
  author = {Itano, Wayne M. and Heinzen, D. J. and Bollinger, J. J. and Wineland, D. J.},
  journal = {Phys. Rev. A},
  volume = {41},
  issue = {5},
  pages = {2295--2300},
  numpages = {0},
  year = {1990},
  month = {Mar},
  publisher = {American Physical Society},
  doi = {10.1103/PhysRevA.41.2295},
  url = {https://link.aps.org/doi/10.1103/PhysRevA.41.2295}
}

@article{Home1986,
doi = {10.1088/0305-4470/19/10/025},
url = {https://doi.org/10.1088/0305-4470/19/10/025},
year = {1986},
month = {jul},
publisher = {},
volume = {19},
number = {10},
pages = {1847},
author = {D Home and M A B Whitaker},
title = {Reflections on the quantum Zeno paradox},
journal = {Journal of Physics A: Mathematical and General},
}

@article{Home1992,
doi = {10.1088/0305-4470/25/3/022},
url = {https://doi.org/10.1088/0305-4470/25/3/022},
year = {1992},
month = {feb},
publisher = {},
volume = {25},
number = {3},
pages = {657},
author = {D Home and M A B Whitaker},
title = {A critical re-examination of the quantum Zeno paradox},
journal = {Journal of Physics A: Mathematical and General}
}

@article{Home1997,
title = {A Conceptual Analysis of Quantum Zeno; Paradox, Measurement, and Experiment},
journal = {Annals of Physics},
volume = {258},
number = {2},
pages = {237-285},
year = {1997},
issn = {0003-4916},
doi = {https://doi.org/10.1006/aphy.1997.5699},
url = {https://www.sciencedirect.com/science/article/pii/S0003491697956992},
author = {D Home and M.A.B Whitaker}
}

@article{Facchi2004,
  title = {Unification of dynamical decoupling and the quantum Zeno effect},
  author = {Facchi, P. and Lidar, D. A. and Pascazio, S.},
  journal = {Phys. Rev. A},
  volume = {69},
  issue = {3},
  pages = {032314},
  numpages = {6},
  year = {2004},
  month = {Mar},
  publisher = {American Physical Society},
  doi = {10.1103/PhysRevA.69.032314},
  url = {https://link.aps.org/doi/10.1103/PhysRevA.69.032314}
}

@article{Georgescu2022,
  author    = {Georgescu, Iulia},
  title     = {Quantum Zeno Effect at 45},
  journal   = {Nature Reviews Physics},
  volume    = {4},
  number    = {5},
  pages     = {289},
  year      = {2022},
  doi       = {10.1038/s42254-022-00454-2}
}

@article{Facchi2002,
  title = {Quantum Zeno Subspaces},
  author = {Facchi, P. and Pascazio, S.},
  journal = {Phys. Rev. Lett.},
  volume = {89},
  issue = {8},
  pages = {080401},
  numpages = {4},
  year = {2002},
  month = {Aug},
  publisher = {American Physical Society},
  doi = {10.1103/PhysRevLett.89.080401},
  url = {https://link.aps.org/doi/10.1103/PhysRevLett.89.080401}
}

@article{Kwiat1995,
  title = {Interaction-Free Measurement},
  author = {Kwiat, Paul and Weinfurter, Harald and Herzog, Thomas and Zeilinger, Anton and Kasevich, Mark A.},
  journal = {Phys. Rev. Lett.},
  volume = {74},
  issue = {24},
  pages = {4763--4766},
  numpages = {0},
  year = {1995},
  month = {Jun},
  publisher = {American Physical Society},
  doi = {10.1103/PhysRevLett.74.4763},
  url = {https://link.aps.org/doi/10.1103/PhysRevLett.74.4763}
}

@article{Signoles2014,
  author    = {Signoles, Adrien and Facon, Adrien and Grosso, Dorian and Dotsenko, Igor and Haroche, Serge and Raimond, Jean-Michel and Brune, Michel and Gleyzes, S{\'e}bastien},
  title     = {Confined Quantum Zeno Dynamics of a Watched Atomic Arrow},
  journal   = {Nature Physics},
  volume    = {10},
  number    = {10},
  pages     = {715--719},
  year      = {2014},
  doi       = {10.1038/nphys3076}
}

@article{Barone2004,
  title = {Dynamical Control of Macroscopic Quantum Tunneling},
  author = {Barone, A. and Kurizki, G. and Kofman, A. G.},
  journal = {Phys. Rev. Lett.},
  volume = {92},
  issue = {20},
  pages = {200403},
  numpages = {4},
  year = {2004},
  month = {May},
  publisher = {American Physical Society},
  doi = {10.1103/PhysRevLett.92.200403},
  url = {https://link.aps.org/doi/10.1103/PhysRevLett.92.200403}
}

@article{Ghasemi2019,
  title = {A New Approach to Study the Zeno Effect for a Macroscopic Quantum System under Frequent Interactions with a Harmonic Environment},
  author = {Ghasemi, F. and Shafiee, A.},
  journal = {Scientific Reports},
  volume = {9},
  pages = {15265},
  year = {2019},
  month = {October},
  publisher = {Nature Publishing Group},
  doi = {10.1038/s41598-019-51729-1},
  url = {https://doi.org/10.1038/s41598-019-51729-1}
}

@article{Shchesnovich2010,
  title = {Control of a Bose-Einstein condensate by dissipation: Nonlinear Zeno effect},
  author = {Shchesnovich, V. S. and Konotop, V. V.},
  journal = {Phys. Rev. A},
  volume = {81},
  issue = {5},
  pages = {053611},
  numpages = {5},
  year = {2010},
  month = {May},
  publisher = {American Physical Society},
  doi = {10.1103/PhysRevA.81.053611},
  url = {https://link.aps.org/doi/10.1103/PhysRevA.81.053611}
}

@article{Barontini2013,
  title = {Controlling the Dynamics of an Open Many-Body Quantum System with Localized Dissipation},
  author = {Barontini, G. and Labouvie, R. and Stubenrauch, F. and Vogler, A. and Guarrera, V. and Ott, H.},
  journal = {Phys. Rev. Lett.},
  volume = {110},
  issue = {3},
  pages = {035302},
  numpages = {5},
  year = {2013},
  month = {Jan},
  publisher = {American Physical Society},
  doi = {10.1103/PhysRevLett.110.035302},
  url = {https://link.aps.org/doi/10.1103/PhysRevLett.110.035302}
}

@article{Fein2019,
  author    = {Fein, Yair Y. and Geyer, Philipp and Zwick, Patrick and Kialka, Filip and Pedalino, Sara and Mayor, Marcel and Gerlich, Stefan and Arndt, Markus},
  title     = {Quantum Superposition of Molecules beyond 25 kDa},
  journal   = {Nature Physics},
  volume    = {15},
  number    = {12},
  pages     = {1242--1245},
  year      = {2019},
  month     = {December},
  doi       = {10.1038/s41567-019-0663-9},
  url       = {https://doi.org/10.1038/s41567-019-0663-9}
}

@article{Bild2023,
author = {Marius Bild  and Matteo Fadel  and Yu Yang  and Uwe von Lüpke  and Phillip Martin  and Alessandro Bruno  and Yiwen Chu },
title = {Schrödinger cat states of a 16-microgram mechanical oscillator},
journal = {Science},
volume = {380},
number = {6642},
pages = {274-278},
year = {2023},
doi = {10.1126/science.adf7553},
URL = {https://www.science.org/doi/abs/10.1126/science.adf7553},
eprint = {https://www.science.org/doi/pdf/10.1126/science.adf7553}
}

@article{MassiveQMreview2025,
  title = {Massive quantum systems as interfaces of quantum mechanics and gravity},
  author = {Bose, Sougato and Fuentes, Ivette and Geraci, Andrew A. and Khan, Saba Mehsar and Qvarfort, Sofia and Rademacher, Markus and Rashid, Muddassar and Toro\ifmmode \check{s}\else \v{s}\fi{}, Marko and Ulbricht, Hendrik and Wanjura, Clara C.},
  journal = {Rev. Mod. Phys.},
  volume = {97},
  issue = {1},
  pages = {015003},
  numpages = {71},
  year = {2025},
  month = {Feb},
  publisher = {American Physical Society},
  doi = {10.1103/RevModPhys.97.015003},
  url = {https://link.aps.org/doi/10.1103/RevModPhys.97.015003}
}

@article{Pedalino2026,
  author    = {Pedalino, S. and Ram{\'\i}rez-Galindo, B. E. and Ferstl, R. and et al.},
  title     = {Probing quantum mechanics with nanoparticle matter-wave interferometry},
  journal   = {Nature},
  volume    = {649},
  pages     = {866--870},
  year      = {2026},
  doi       = {10.1038/s41586-025-09917-9},
  url       = {https://doi.org/10.1038/s41586-025-09917-9}
}

@article{DasMassindependent2024,
  title = {Mass-Independent Scheme to Test the Quantumness of a Massive Object},
  author = {Das, Debarshi and Home, Dipankar and Ulbricht, Hendrik and Bose, Sougato},
  journal = {Phys. Rev. Lett.},
  volume = {132},
  issue = {3},
  pages = {030202},
  numpages = {7},
  year = {2024},
  month = {Jan},
  publisher = {American Physical Society},
  doi = {10.1103/PhysRevLett.132.030202},
  url = {https://link.aps.org/doi/10.1103/PhysRevLett.132.030202}
}

@article{Kofler2013,
  title = {Condition for macroscopic realism beyond the Leggett-Garg inequalities},
  author = {Kofler, Johannes and Brukner, Caslav},
  journal = {Phys. Rev. A},
  volume = {87},
  issue = {5},
  pages = {052115},
  numpages = {5},
  year = {2013},
  month = {May},
  publisher = {American Physical Society},
  doi = {10.1103/PhysRevA.87.052115},
  url = {https://link.aps.org/doi/10.1103/PhysRevA.87.052115}
}

@article{Schild2015,
  title = {Maximum violations of the quantum-witness equality},
  author = {Schild, Greg and Emary, Clive},
  journal = {Phys. Rev. A},
  volume = {92},
  issue = {3},
  pages = {032101},
  numpages = {6},
  year = {2015},
  month = {Sep},
  publisher = {American Physical Society},
  doi = {10.1103/PhysRevA.92.032101},
  url = {https://link.aps.org/doi/10.1103/PhysRevA.92.032101}
}

@article{Knee2016,
  author    = {Knee, George C. and Kakuyanagi, Kosuke and Yeh, Mao-Chuang and Matsuzaki, Yuichiro and Toida, Hiraku and Yamaguchi, Hiroshi and Saito, Shiro and Leggett, Anthony J. and Munro, William J.},
  title     = {A Strict Experimental Test of Macroscopic Realism in a Superconducting Flux Qubit},
  journal   = {Nature Communications},
  volume    = {7},
  pages     = {13253},
  year      = {2016},
  doi       = {10.1038/ncomms13253}
}

@article{Aspelmeyer2014,
  title = {Cavity optomechanics},
  author = {Aspelmeyer, Markus and Kippenberg, Tobias J. and Marquardt, Florian},
  journal = {Rev. Mod. Phys.},
  volume = {86},
  issue = {4},
  pages = {1391--1452},
  numpages = {62},
  year = {2014},
  month = {Dec},
  publisher = {American Physical Society},
  doi = {10.1103/RevModPhys.86.1391},
  url = {https://link.aps.org/doi/10.1103/RevModPhys.86.1391}
}

@article{deLosRiosSommer2021,
  author    = {de los R{\'i}os Sommer, Andr{\'e}s and Meyer, Nadine and Quidant, Romain},
  title     = {Strong Optomechanical Coupling at Room Temperature by Coherent Scattering},
  journal   = {Nature Communications},
  volume    = {12},
  number    = {1},
  pages     = {276},
  year      = {2021},
  doi       = {10.1038/s41467-020-20419-2}
}

@article{Groblacher2009,
  author    = {Gr{\"o}blacher, Simon and Hammerer, Klemens and Vanner, Michael R. and Aspelmeyer, Markus},
  title     = {Observation of Strong Coupling Between a Micromechanical Resonator and an Optical Cavity Field},
  journal   = {Nature},
  volume    = {460},
  number    = {7256},
  pages     = {724--727},
  year      = {2009},
  doi       = {10.1038/nature08171}
}

@article{HanifTesting2024,
  title = {Testing Whether Gravity Acts as a Quantum Entity When Measured},
  author = {Hanif, Farhan and Das, Debarshi and Halliwell, Jonathan and Home, Dipankar and Mazumdar, Anupam and Ulbricht, Hendrik and Bose, Sougato},
  journal = {Phys. Rev. Lett.},
  volume = {133},
  issue = {18},
  pages = {180201},
  numpages = {7},
  year = {2024},
  month = {Oct},
  publisher = {American Physical Society},
  doi = {10.1103/PhysRevLett.133.180201},
  url = {https://link.aps.org/doi/10.1103/PhysRevLett.133.180201}
}

@misc{bosespin2025,
      title={A Spin-Based Pathway to Testing the Quantum Nature of Gravity}, 
      author={Sougato Bose and Anupam Mazumdar and Roger Penrose and Ivette Fuentes and Marko Toroš and Ron Folman and Gerard J. Milburn and Myungshik Kim and Adrian Kent and A. T. M. Anishur Rahman and Cyril Laplane and Aaron Markowitz and Debarshi Das and Ethan Campos-Méndez and Eva Kilian and David Groswasser and Menachem Givon and Or Dobkowski and Peter Skakunenko and Maria Muretova and Yonathan Japha and Naor Levi and Omer Feldman and Damián Pitalúa-García and Jonathan M. H. Gosling and Ka-Di Zhu and Marco Genovese and Kia Romero-Hojjati and Ryan J. Marshman and Markus Rademacher and Martine Schut and Melanie Bautista-Cruz and Qian Xiang and Stuart M. Graham and James E. March and William J. Fairbairn and Karishma S. Gokani and Joseph Aziz and Richard Howl and Run Zhou and Ryan Rizaldy and Thiago Guerreiro and Tian Zhou and Jason Twamley and Chiara Marletto and Vlatko Vedral and Jonathan Oppenheim and Mauro Paternostro and Hendrik Ulbricht and Peter F. Barker and Thomas P. Purdy and M. V. Gurudev Dutt and Andrew A. Geraci and David C. Moore and Gavin W. Morley},
      year={2025},
      eprint={2509.01586},
      archivePrefix={arXiv},
      primaryClass={quant-ph},
      url={https://arxiv.org/abs/2509.01586}, 
}

@inproceedings{Itano1997,
author = {Wayne M. Itano and Christopher R. Monroe and D. M. Meekhof and D. Leibfried and B. E. King and David J. Wineland},
title = {{Quantum harmonic oscillator state synthesis and analysis}},
volume = {2995},
booktitle = {Atom Optics},
editor = {Mara Goff Prentiss and William D. Phillips},
organization = {International Society for Optics and Photonics},
publisher = {SPIE},
pages = {43 -- 55},
year = {1997},
doi = {10.1117/12.273771},
URL = {https://doi.org/10.1117/12.273771}
}

@misc{zindorfHow2025,
      title={How "Quantum" is your Quantum Computer? Macrorealism-based Benchmarking via Mid-Circuit Parity Measurements}, 
      author={Ben Zindorf and Lorenzo Braccini and Debarshi Das and Sougato Bose},
      year={2025},
      eprint={2511.15881},
      archivePrefix={arXiv},
      primaryClass={quant-ph},
      url={https://arxiv.org/abs/2511.15881}, 
}

@misc{bracciniNonclassicality2025,
      title={Nonclassicality of a Macroscopic Qubit-Ensemble via Parity Measurement Induced Disturbance}, 
      author={Lorenzo Braccini and Debarshi Das and Ben Zindorf and Stephen D. Hogan and John J. L. Morton and Sougato Bose},
      year={2025},
      eprint={2511.15880},
      archivePrefix={arXiv},
      primaryClass={quant-ph},
      url={https://arxiv.org/abs/2511.15880}, 
}

@article{BurgarthExponential2014,
  author    = {Daniel Klaus Burgarth and Paolo Facchi and Vittorio Giovannetti and Hiromichi Nakazato and Saverio Pascazio and Kazuya Yuasa},
  title     = {Exponential rise of dynamical complexity in quantum computing through projections},
  journal   = {Nature Communications},
  volume    = {5},
  pages      = {5173},
  year       = {2014},
  doi        = {10.1038/ncomms6173},
  url        = {https://doi.org/10.1038/ncomms6173},
  publisher  = {Nature Publishing Group}
}

@article{Diosi1987,
title = {A universal master equation for the gravitational violation of quantum mechanics},
journal = {Physics Letters A},
volume = {120},
number = {8},
pages = {377-381},
year = {1987},
issn = {0375-9601},
doi = {https://doi.org/10.1016/0375-9601(87)90681-5},
url = {https://www.sciencedirect.com/science/article/pii/0375960187906815},
author = {L. Diósi}
}

@article{Penrose1996,
  author    = {Roger Penrose},
  title     = {On Gravity's Role in Quantum State Reduction},
  journal   = {General Relativity and Gravitation},
  volume    = {28},
  number    = {5},
  pages     = {581--600},
  year      = {1996},
  month     = {May},
  doi       = {10.1007/BF02105068},
  url       = {https://doi.org/10.1007/BF02105068}
}

@article{Ballestero2021,
author = {C. Gonzalez-Ballestero  and M. Aspelmeyer  and L. Novotny  and R. Quidant  and O. Romero-Isart },
title = {Levitodynamics: Levitation and control of microscopic objects in vacuum},
journal = {Science},
volume = {374},
number = {6564},
pages = {eabg3027},
year = {2021},
doi = {10.1126/science.abg3027},
URL = {https://www.science.org/doi/abs/10.1126/science.abg3027}
}

@article{Gieseler2012,
  title = {Subkelvin Parametric Feedback Cooling of a Laser-Trapped Nanoparticle},
  author = {Gieseler, Jan and Deutsch, Bradley and Quidant, Romain and Novotny, Lukas},
  journal = {Phys. Rev. Lett.},
  volume = {109},
  issue = {10},
  pages = {103603},
  numpages = {5},
  year = {2012},
  month = {Sep},
  publisher = {American Physical Society},
  doi = {10.1103/PhysRevLett.109.103603},
  url = {https://link.aps.org/doi/10.1103/PhysRevLett.109.103603}
}

@article{Doherty1999,
  title = {Feedback control of quantum systems using continuous state estimation},
  author = {Doherty, A. C. and Jacobs, K.},
  journal = {Phys. Rev. A},
  volume = {60},
  issue = {4},
  pages = {2700--2711},
  numpages = {0},
  year = {1999},
  month = {Oct},
  publisher = {American Physical Society},
  doi = {10.1103/PhysRevA.60.2700},
  url = {https://link.aps.org/doi/10.1103/PhysRevA.60.2700}
}

@article{Walker2019,
  title = {Measurement and feedback for cooling heavy levitated particles in low-frequency traps},
  author = {Walker, L. S. and Robb, G. R. M. and Daley, A. J.},
  journal = {Phys. Rev. A},
  volume = {100},
  issue = {6},
  pages = {063819},
  numpages = {9},
  year = {2019},
  month = {Dec},
  publisher = {American Physical Society},
  doi = {10.1103/PhysRevA.100.063819},
  url = {https://link.aps.org/doi/10.1103/PhysRevA.100.063819}
}

@article{Vinante2019,
  title = {Testing collapse models with levitated nanoparticles: Detection challenge},
  author = {Vinante, A. and Pontin, A. and Rashid, M. and Toro\ifmmode \check{s}\else \v{s}\fi{}, M. and Barker, P. F. and Ulbricht, H.},
  journal = {Phys. Rev. A},
  volume = {100},
  issue = {1},
  pages = {012119},
  numpages = {13},
  year = {2019},
  month = {Jul},
  publisher = {American Physical Society},
  doi = {10.1103/PhysRevA.100.012119},
  url = {https://link.aps.org/doi/10.1103/PhysRevA.100.012119}
}

@article{Donadi2021,
  author  = {Donadi, Sandro and Piscicchia, Kristian and Curceanu, Catalina and others},
  title   = {Underground test of gravity-related wave function collapse},
  journal = {Nature Physics},
  volume  = {17},
  pages   = {74--78},
  year    = {2021},
  doi     = {10.1038/s41567-020-1008-4}
}

@article{Montenegro2022,
  title = {Sequential Measurements for Quantum-Enhanced Magnetometry in Spin Chain Probes},
  author = {Montenegro, Victor and Jones, Gareth Si\^on and Bose, Sougato and Bayat, Abolfazl},
  journal = {Phys. Rev. Lett.},
  volume = {129},
  issue = {12},
  pages = {120503},
  numpages = {7},
  year = {2022},
  month = {Sep},
  publisher = {American Physical Society},
  doi = {10.1103/PhysRevLett.129.120503},
  url = {https://link.aps.org/doi/10.1103/PhysRevLett.129.120503}
}

@article{Yang2023,
  title = {Extractable information capacity in sequential measurements metrology},
  author = {Yang, Yaoling and Montenegro, Victor and Bayat, Abolfazl},
  journal = {Phys. Rev. Res.},
  volume = {5},
  issue = {4},
  pages = {043273},
  numpages = {12},
  year = {2023},
  month = {Dec},
  publisher = {American Physical Society},
  doi = {10.1103/PhysRevResearch.5.043273},
  url = {https://link.aps.org/doi/10.1103/PhysRevResearch.5.043273}
}

@misc{Das2026How,
      title={How to Test Bell Nonlocality for Gravity?}, 
      author={Debarshi Das and Mir Alimuddin and Simon Storz and Yiwen Chu and Sougato Bose},
      year={2026},
      eprint={2608.08778},
      archivePrefix={arXiv},
      primaryClass={quant-ph},
      url={https://arxiv.org/abs/2608.08778}, 
}


\onecolumngrid
\renewcommand{\theequation}{S\arabic{equation}}
\setcounter{equation}{0}
\noindent \begin{center}{\Large \bf Supplemental Material}\end{center}
~\vspace{-0.5cm}


\section{Sec.~I: Analytical derivation of the upper and lower bounds for $\kappa_{\mathrm{NDC}}$}\label{SecI}

In this section, we derive the analytical lower and upper bounds on the non-disturbance condition (NDC) violation parameter $\kappa_{\mathrm{NDC}}$ for a coherently evolving mechanical harmonic oscillator subjected to repeated projective measurements. The derivation presented here complements the analysis discussed in the main text and explicitly demonstrates how repeated measurements lead to the quantum Zeno enhancement of measurement-induced nonclassicality.

We consider a mechanical harmonic oscillator of angular frequency $\Omega_m$, initially prepared in the coherent state
$$
\rho_0 = |\alpha\rangle_m\langle\alpha|.
$$

The system evolves under the Hamiltonian
$$
\hat{H}
=
\hbar\Omega_m
\left(
\hat{n}
+
\frac{\mathbb{I}}{2}
\right).
$$

At time $t=\delta t$, we perform the two-outcome measurement
$$
\left\{
E_0
=
|\alpha\rangle_m\langle\alpha|,
\quad
E_1
=
\mathbb{I}_m
-
|\alpha\rangle_m\langle\alpha|
\right\},
$$
where the outcomes associated with $E_0$ and $E_1$ are denoted by $0$ and $1$, respectively. The same measurement is repeated after every interval $\delta t$. After $(N-1)$ intermediate measurements, the final measurement is performed at
$$
t_f=N\delta t.
$$

The probability of obtaining the outcome $0$ in the final measurement, while summing over all possible intermediate outcomes, is given by
\begin{equation*}
P^{(\mathrm{with})}_0(t_f)
=
\sum_{a_1,\dots,a_{N-1}=0,1}
P(a_1,a_2,\dots,a_{N-1},a_N=0),
\label{eq:S1}
\end{equation*}
where $P(a_1,a_2,\dots,a_N)$ denotes the joint probability associated with a complete sequence of measurement outcomes.

In the absence of intermediate measurements, the evolved state at time $t_f$ becomes
\begin{equation*}
|\psi(t_f)\rangle
=
e^{-i\hat{H}t_f/\hbar}
|\alpha\rangle
=
e^{-i\Omega_m t_f/2}
|\alpha e^{-i\Omega_m t_f}\rangle .
\label{eq:S2}
\end{equation*}

Using the overlap relation between coherent states,
$$
\langle\beta|\gamma\rangle
=
\exp
\left[
-
\frac{|\beta|^2}{2}
-
\frac{|\gamma|^2}{2}
+
\beta^*\gamma
\right],
$$
the probability of obtaining the outcome $0$ in the final measurement becomes
\begin{align}
P^{(\mathrm{without})}_0(t_f)
&=
\left|
\langle\alpha|
\alpha e^{-i\Omega_m t_f}
\rangle
\right|^2
\nonumber\\
&=
\left|
\exp
\left[
-
\frac{1}{2}
\left(
|\alpha|^2
+
|\alpha|^2
-
2|\alpha|^2 e^{-i\Omega_m t_f}
\right)
\right]
\right|^2
\nonumber\\
&=
\exp
\left[
-
2|\alpha|^2
\left(
1-\cos\Omega_m t_f
\right)
\right].
\label{eq:S3}
\end{align}

The no-disturbance condition introduced in the main text is
\begin{equation}
\kappa_{\mathrm{NDC}}
=
P^{(\mathrm{with})}_0(t_f)
-
P^{(\mathrm{without})}_0(t_f).
\label{eq:S4}
\end{equation}

For any classical noninvasive description, $\kappa_{\mathrm{NDC}}=0.$

Hence, any nonzero value of $\kappa_{\mathrm{NDC}}$ signals measurement-induced disturbance and therefore nonclassicality.

Throughout the analysis, we fix the total evolution phase as
\begin{equation}
\Omega_m t_f
=
\Omega_m N\delta t
\equiv
\theta,
\label{eq:S5}
\end{equation}
where $\theta$ is finite and independent of $N$. Consequently, increasing the number of measurements requires decreasing $\delta t$ such that Eq.~\eqref{eq:S5} remains satisfied.

To obtain an analytical lower bound on $P^{(\mathrm{with})}_0(t_f),$
we first note that
\begin{align}
P^{(\mathrm{with})}_0(t_f)
&=
\sum_{a_1,\dots,a_{N-1}=0,1}
P(a_1,a_2,\dots,a_{N-1},a_N=0)
\nonumber\\
&\geq
P(a_1=0,a_2=0,\dots,a_N=0),
\label{eq:S6}
\end{align}
where the right-hand side corresponds to the specific trajectory in which every measurement yields the outcome $0$.

We now evaluate this contribution explicitly. Between two successive measurements separated by $\delta t$, the coherent state evolves as
$$
|\alpha\rangle
\rightarrow
|\alpha e^{-i\Omega_m\delta t}\rangle.
$$

Therefore, the probability of obtaining the outcome $0$ in a single measurement step is
\begin{align}
p_0(\delta t)
&=
\left|
\langle\alpha|
\alpha e^{-i\Omega_m\delta t}
\rangle
\right|^2
\nonumber\\
&=
\exp
\left[
-
2|\alpha|^2
\left(
1-\cos\Omega_m\delta t
\right)
\right].
\label{eq:S7}
\end{align}

Since the state is projected back onto $|\alpha\rangle$ after each successful outcome $0$, the probability of obtaining the outcome $0$ in all $N$ measurements is simply
\begin{equation}
P(a_1=0,\dots,a_N=0)
=
\left[
p_0(\delta t)
\right]^N .
\label{eq:S8}
\end{equation}

Substituting Eq.~\eqref{eq:S7} into Eq.~\eqref{eq:S8} gives
\begin{equation}
P(a_1=0,\dots,a_N=0)
=
\left[
\exp
\left(
-
2|\alpha|^2
\left[
1-\cos(\Omega_m\delta t)
\right]
\right)
\right]^N .
\label{eq:S9}
\end{equation}

We now consider the short-time regime $\Omega_m\delta t\ll 1.$

Expanding the cosine function up to second order,
\begin{equation}
\cos(\Omega_m\delta t)
\simeq
1
-
\frac{\Omega_m^2(\delta t)^2}{2},
\label{eq:S10}
\end{equation}
where terms of order $\Omega_m^4(\delta t)^4$ have been neglected.

Using Eq.~\eqref{eq:S10} in Eq.~\eqref{eq:S9}, we obtain
\begin{align}
P(a_1=0,\dots,a_N=0)
&\simeq
\exp
\left[
-
N|\alpha|^2
\Omega_m^2(\delta t)^2
\right]
\nonumber\\
&=
\exp
\left[
-
\frac{|\alpha|^2\Omega_m^2 t_f^2}{N}
\right]
\nonumber\\
&=
\exp
\left[
-
\frac{|\alpha|^2\theta^2}{N}
\right],
\label{eq:S11}
\end{align}
where we have used $t_f=N\delta t$ together with Eq.~\eqref{eq:S5}.

Combining Eqs.~\eqref{eq:S6} and \eqref{eq:S11}, we obtain the lower bound
\begin{equation}
P^{(\mathrm{with})}_0(t_f)
\geq
\exp
\left[
-
\frac{|\alpha|^2\theta^2}{N}
\right].
\label{eq:S12}
\end{equation}

Substituting Eqs.~\eqref{eq:S3} and \eqref{eq:S12} into Eq.~\eqref{eq:S4}, we obtain
\begin{equation}
\kappa_{\mathrm{NDC}}
\geq
\exp
\left[
-
\frac{|\alpha|^2\theta^2}{N}
\right]
-
\exp
\left[
-
2|\alpha|^2
(1-\cos\theta)
\right].
\label{eq:S13}
\end{equation}

To analyze the large-$N$ limit, we assume that $1-\cos\theta$ remains finite and that the coherent-state amplitude scales sublinearly with $N$ according to
\begin{equation}
|\alpha|
=
cN^{1/x},
\qquad
x>2,
\label{eq:S14}
\end{equation}
where $c>0$ is finite.

Substituting Eq.~\eqref{eq:S14} into Eq.~\eqref{eq:S12} yields
\begin{equation*}
P^{(\mathrm{with})}_0(t_f)
\geq
\exp
\left[
-
\frac{c^2\theta^2}{N^{1-2/x}}
\right].
\label{eq:S15}
\end{equation*}

Since
$$
1-\frac{2}{x}>0,
$$
we obtain
\begin{equation*}
\lim_{N\rightarrow\infty}
\exp
\left[
-
\frac{c^2\theta^2}{N^{1-2/x}}
\right]
=
1.
\label{eq:S16}
\end{equation*}

Using the normalization of probabilities,
\begin{equation}
\lim_{N\rightarrow\infty}
P^{(\mathrm{with})}_0(t_f)
=
1.
\label{eq:S17}
\end{equation}

On the other hand,
\begin{align}
\lim_{N\rightarrow\infty}
P^{(\mathrm{without})}_0(t_f)
&=
\lim_{N\rightarrow\infty}
\exp
\left[
-
2c^2N^{2/x}
(1-\cos\theta)
\right]
\nonumber\\
&=
0.
\label{eq:S18}
\end{align}

Consequently,
\begin{equation}
\lim_{N\rightarrow\infty}
\kappa_{\mathrm{NDC}}
=
1.
\label{eq:S19}
\end{equation}

Equations~\eqref{eq:S17}-\eqref{eq:S19} demonstrate that, in the limit of infinitely frequent measurements, the repeated projections inhibit the free harmonic evolution and effectively freeze the oscillator in its initial coherent state. In contrast, when no intermediate measurements are performed, the survival probability vanishes. This behavior constitutes the quantum Zeno effect. Simultaneously, the NDC violation approaches its algebraic maximum, demonstrating the amplification of measurement-induced nonclassicality through repeated measurements.

For finite $N$, the exact analytical evaluation of $
\kappa_{\mathrm{NDC}}$ is generally nontrivial because $P^{(\mathrm{with})}_0(t_f)$ contains contributions from all possible intermediate measurement histories. Nevertheless, one can derive rigorous lower and upper bounds.

Defining
$$
L_{\mathrm{NDC}}
\leq
\kappa_{\mathrm{NDC}}
\leq
U_{\mathrm{NDC}},
$$
the lower bound follows directly from Eq.~\eqref{eq:S13},
\begin{equation}
L_{\mathrm{NDC}}
=
\exp
\left[
-
\frac{|\alpha|^2\theta^2}{N}
\right]
-
\exp
\left[
-
2|\alpha|^2
(1-\cos\theta)
\right].
\label{eq:S20}
\end{equation}

The upper bound follows from the trivial inequality
$$
P^{(\mathrm{with})}_0(t_f)\leq 1,
$$
which gives
\begin{equation}
U_{\mathrm{NDC}}
=
1
-
\exp
\left[
-
2|\alpha|^2
(1-\cos\theta)
\right].
\label{eq:S21}
\end{equation}

Therefore, for finite $N$, the exact value of $\kappa_{\mathrm{NDC}}$ lies between the analytical bounds in Eqs.~\eqref{eq:S20} and \eqref{eq:S21}. The quantitative evaluation of the finite-measurement dynamics is presented numerically in Sec.~II.
\section{Sec.~II: Numerical evaluation of $\kappa_{\mathrm{NDC}}$}
\label{SecII}
In this section, we present the numerical framework employed to evaluate 
$\kappa_{\mathrm{NDC}}$. Since the dynamics are formulated in the Fock basis, 
the infinite-dimensional Hilbert space associated with the harmonic oscillator 
must be truncated to a finite dimension suitable for numerical computation. 
We therefore begin by constructing the truncated coherent state representation, 
followed by the corresponding finite-dimensional Hamiltonian and the iterative 
measurement-evolution protocol used throughout the simulations.

A coherent state $|\alpha\rangle$  is defined as an eigenstate of the annihilation operator $\hat a$,
\begin{equation*}
\hat a |\alpha\rangle = \alpha |\alpha\rangle,
\end{equation*}
where $\alpha \in \mathbb{C}$ denotes the coherent-state amplitude. 
In the Fock basis, the coherent state admits the expansion
\begin{equation*}
|\alpha\rangle
=
e^{-|\alpha|^2/2}
\sum_{n=0}^{\infty}
\frac{\alpha^n}{\sqrt{n!}}
|n\rangle .
\label{eq:coherent_expansion}
\end{equation*}

The corresponding occupation-number statistics follow a Poissonian distribution,
\begin{equation*}
P(n,\alpha)
=
e^{-|\alpha|^2}
\frac{|\alpha|^{2n}}{n!},
\label{poisson_distribution}
\end{equation*}
whose mean occupation number is $\langle n\rangle = |\alpha|^2 $.

Since the probability distribution decays exponentially for sufficiently large 
$n$, numerical calculations can be accurately performed within a truncated 
Hilbert space of finite dimension $n_{\max}+1$. The cutoff $n_{\max}$ is chosen such that the neglected probability weight remains below a prescribed threshold $\varepsilon$,
\begin{equation*}
P(n_{\max},\alpha)<\varepsilon .
\end{equation*}

Consequently, the coherent state is approximated as
\begin{equation*}
|\alpha\rangle_{\mathrm{tr}}
=
\mathcal{N}_{\alpha}
\sum_{n=0}^{n_{\max}}
\frac{\alpha^n}{\sqrt{n!}}
|n\rangle ,
\label{truncated_coherent}
\end{equation*}
where
\begin{equation*}
\mathcal{N}_{\alpha}
=
\left(
\sum_{n=0}^{n_{\max}}
\frac{|\alpha|^{2n}}{n!}
\right)^{-1/2}
\end{equation*}
is the normalization factor associated with the truncated state.

For sufficiently large mean occupation numbers 
($|\alpha|^2 \gtrsim 10$), the Poissonian distribution approaches a Gaussian distribution according to the central-limit approximation,
\begin{equation*}
P(n,\alpha)
\sim
\mathcal{G}
\left(
\mu = |\alpha|^2,
\sigma = \sqrt{|\alpha|^2}
\right),
\end{equation*}
where $\mu$ and $\sigma$ denote the mean and standard deviation, respectively. 
This approximation provides an efficient estimate of the relevant Fock-space support prior to the numerical truncation.

\subsection{Determination of the cutoff dimension}

The cutoff \(n_{\max}\) is determined iteratively by monitoring the tail of the occupation-number distribution until the corresponding probability falls below the chosen numerical threshold \(\varepsilon\). The procedure used in the simulations is summarized below.

\textbf{Procedure for determining the Fock-space cutoff $n_{\max}$:}

\begin{enumerate}
    \item For a coherent state with amplitude $\alpha$, define the photon-number distribution
    \begin{align*}
    P(n,\alpha)
    =
    e^{-|\alpha|^2}
    \frac{|\alpha|^{2n}}{n!}.
    \end{align*}

    \item Choose a threshold probability $\varepsilon$ and a search parameter $k$.

    \item Set the search limit as
    \begin{align*}
    n_{\mathrm{search}}
    =
    \left\lceil
    |\alpha|^2+k|\alpha|
    \right\rceil .
    \end{align*}

    \item Initialize $n_{\max}=0$.

    \item For each integer $n=0,1,\ldots,n_{\mathrm{search}}$, evaluate $P(n,\alpha)$.

    \item If
    \begin{align*}
    P(n,\alpha)\ge\varepsilon,
    \end{align*}
    update
    \begin{align*}
    n_{\max}=n.
    \end{align*}

    \item After all values of $n$ have been checked, take the largest value satisfying the threshold criterion as the Fock-space cutoff and return $n_{\max}$.
\end{enumerate}

Once the cutoff dimension is obtained, all operators and states are projected onto the truncated Hilbert space ($\mathcal{H}_{\mathrm{tr}}$)
\begin{equation*}
\mathcal{H}_{\mathrm{tr}}
=
\mathrm{span}
\left\{
|0\rangle,|1\rangle,\dots,|n_{\max}\rangle
\right\}.
\end{equation*}

\subsection{Truncated harmonic-oscillator Hamiltonian}

The Hamiltonian of the quantum harmonic oscillator is
\begin{equation}
H_m=\hbar\Omega_m
\left(
\hat a^\dagger \hat a + \frac{1}{2}
\right),
\label{eq:harmonic_hamiltonian}
\end{equation}
where $\Omega_m$ is the oscillator frequency.

In the Fock basis, the number operator satisfies
\begin{equation*}
\hat a^\dagger \hat a |n\rangle = n |n\rangle,
\end{equation*}
so that the Hamiltonian acts diagonally as
\begin{equation*}
H_m|n\rangle
=
\hbar\Omega_m
\left(
n+\frac{1}{2}
\right)
|n\rangle .
\end{equation*}

Restricting the dynamics to the truncated Hilbert space 
\(\mathcal{H}_{\mathrm{tr}}\), the Hamiltonian becomes
\begin{equation*}
H_{\mathrm{tr}}
=
\hbar\Omega_m
\sum_{n=0}^{n_{\max}}
\left(
n+\frac{1}{2}
\right)
|n\rangle\langle n|.
\label{eq:truncated_hamiltonian}
\end{equation*}

Correspondingly, the annihilation and creation operators are approximated by
\begin{equation*}
\hat a_{\mathrm{tr}}
=
\sum_{n=1}^{n_{\max}}
\sqrt{n}\,
|n-1\rangle\langle n|,
\end{equation*}
and
\begin{equation*}
\hat a_{\mathrm{tr}}^\dagger
=
\sum_{n=0}^{n_{\max}-1}
\sqrt{n+1}\,
|n+1\rangle\langle n|.
\end{equation*}

The unitary time-evolution operator within the truncated space is therefore $U(\delta t)
=
e^{-i H_{\mathrm{tr}}\delta t/\hbar}.$ Since $H_{\mathrm{tr}}$ is diagonal in the Fock basis, the evolution operator takes the explicit form
\begin{equation*}
U(\delta t)
=
\sum_{n=0}^{n_{\max}}
e^{-i\Omega_m(n+\frac12)\delta t}
|n\rangle\langle n|.
\label{explicit_unitary}
\end{equation*}

\subsection{Iterative measurement-evolution protocol}

After constructing the truncated coherent state and the corresponding finite-dimensional Hamiltonian, we numerically implement the repeated measurement dynamics used to evaluate $\kappa_{\mathrm{NDC}}$.

\noindent\textit{Step 1: Initial state preparation.}
The initial state is represented by the truncated density matrix
\begin{equation}
\rho_0
=
|\alpha\rangle_{\mathrm{tr}}
\langle\alpha|,
\label{eq:rho0}
\end{equation}
defined within the truncated Hilbert space $\mathcal{H}_{\mathrm{tr}}$.

\noindent\textit{Step 2: Unitary evolution.}
The system evolves for a short interval $\delta t$ under the truncated Hamiltonian $H_{\mathrm{tr}}$,
\begin{equation}
\rho_0'
=
U(\delta t)\rho_0 U^\dagger(\delta t),
\label{eq:unitary_evolution}
\end{equation}
where the unitary operator is given by
\begin{equation*}
U(\delta t)
=
e^{-iH_{\mathrm{tr}}\delta t/\hbar}.
\label{unitary_operator_app}
\end{equation*}
Using Eq.~(\ref{eq:rho0}), the evolved state in Eq.~\eqref{eq:unitary_evolution} is evaluated within the truncated Fock space.

\noindent\textit{Step 3: POVM measurement update.}
The two-outcome POVM $\{E_{0}=|\alpha\rangle_m\langle\alpha|, E_{1}=\mathbb{I}_m-|\alpha\rangle_m\langle\alpha|\}$ is applied to the evolved state of Eq.~\eqref{eq:unitary_evolution}. The probability corresponding to the $i$-th outcome is
\begin{equation}
p_{i,t}
=
\Tr[E_i\rho(t)],
\qquad i\in\{0,1\}.
\label{eq:measurement_probability}
\end{equation}

The associated post-measurement state is
\begin{equation}
\rho_i(t)
=
\frac{\sqrt{E_i}\rho(t)\sqrt{E_i}}{p_{i,t}}.
\label{eq:post_measurement_state}
\end{equation}

Since the individual measurement outcomes are not retained, the effective post-measurement state becomes
\begin{equation}
\rho^{(k)}(t)
=
\sum_{i=0}^{1}
p_{i,t}\rho_i(t).
\label{eq:effective_state}
\end{equation}
That is, we consider here the specific Kraus operators $\{K_{\rm no} = \sqrt{E_0}, K_{\rm click} = \sqrt{E_1}\}$.
Here, Eq.~\eqref{eq:measurement_probability} and Eq.~\eqref{eq:post_measurement_state} are used to construct the averaged state in Eq.~\eqref{eq:effective_state}.

\noindent\textit{Step 4: Repeated measurement dynamics.}
The sequence of unitary evolution and POVM update is iterated for $N$ measurement cycles. At the $k$-th step,
\begin{equation}
\rho^{(k-1)}
\longrightarrow
U(\delta t)\rho^{(k-1)}U^\dagger(\delta t)
\longrightarrow
\rho^{(k)},
\label{eq:iterative_dynamics}
\end{equation}
where the update rule follows from Eq.~\eqref{eq:effective_state}.

\noindent\textit{Step 5: Final measurement statistics.}
After the final measurement cycle, the probability associated with the outcome $E_0$ is evaluated as
\begin{equation*}
P_{0}
=
\Tr\left[
E_0
\rho^{(N-1)}(N\delta t)
\right].
\label{eq:final_probability}
\end{equation*}

More generally, the expectation value of an observable $O$ after the \(k\)-th measurement step is given by
\begin{equation}
\langle O\rangle_k
=
\Tr\left[
O\,\rho^{(k)}(k\delta t)
\right].
\label{eq:observable_expectation}
\end{equation}
The state \(\rho^{(k)}\) appearing in Eq.~\eqref{eq:observable_expectation} is obtained iteratively from Eq.~\eqref{eq:iterative_dynamics}.

\begin{table}[t]
\centering
\caption{
Comparison between the numerical and analytical survival probabilities for
different coherent-state amplitudes $\alpha$ and truncation thresholds
$\epsilon$. The corresponding Fock-space cutoff $n_{\max}$ and the final
absolute error $\Delta$ are shown.
}
\label{tab:zeno_error}

\begin{tabular}{cccc}
\toprule
$\alpha$ & $\epsilon$ & $n_{\max}$ & Final $\Delta$ \\
\midrule

0.5 & $10^{-6}$  & 5  & $1.66\times10^{-6}$ \\
0.5 & $10^{-10}$ & 8  & $1.16\times10^{-10}$ \\
0.5 & $10^{-14}$ & 10 & $1.13\times10^{-13}$ \\

\midrule

2.0 & $10^{-6}$  & 16 & $1.66\times10^{-5}$ \\
2.0 & $10^{-10}$ & 22 & $1.74\times10^{-9}$ \\
2.0 & $10^{-14}$ & 27 & $2.35\times10^{-13}$ \\

\midrule

5.0 & $10^{-6}$  & 51 & $1.48\times10^{-6}$ \\
5.0 & $10^{-10}$ & 62 & $1.91\times10^{-10}$ \\
5.0 & $10^{-14}$ & 72 & $8.52\times10^{-15}$ \\

\bottomrule
\end{tabular}

\end{table}

\begin{figure}[t]
\centering
\includegraphics[width=0.5\textwidth]{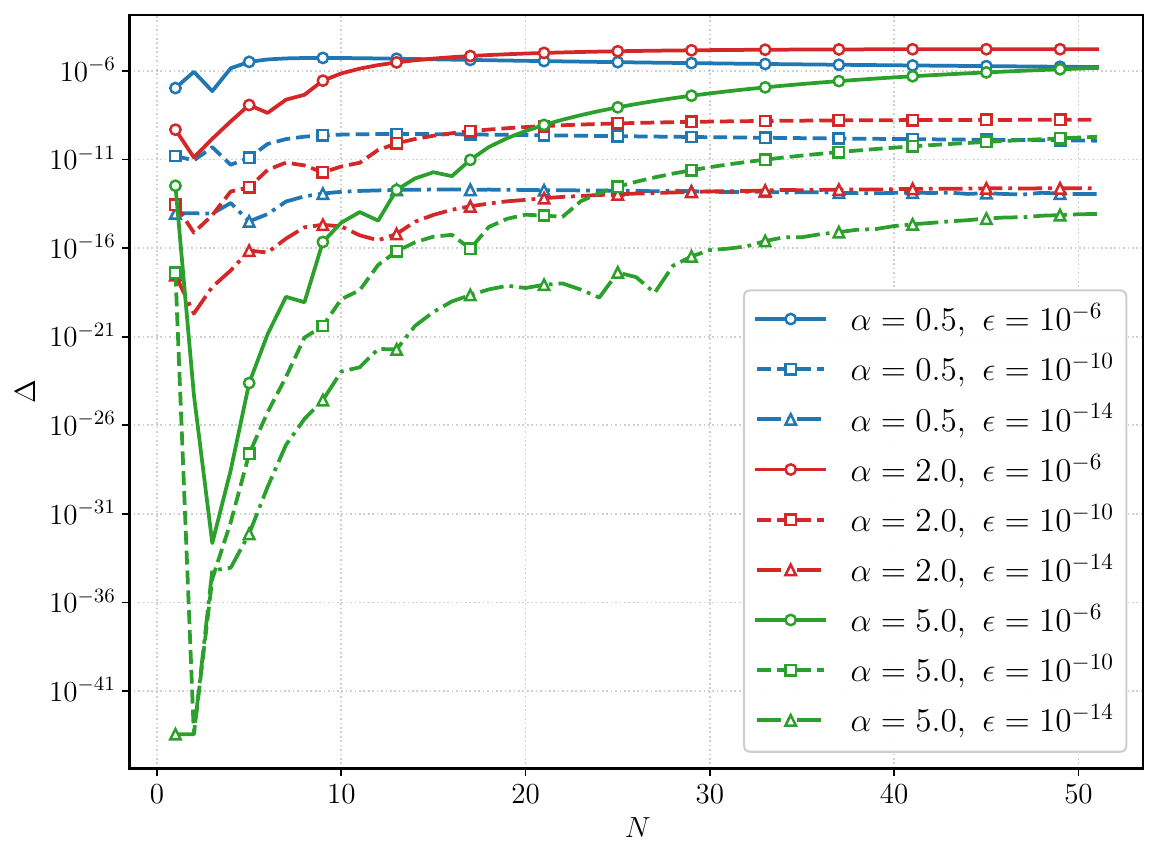}
\caption{
Log-scale comparison between the numerical and analytical survival probabilities for repeated coherent-state measurements under harmonic-oscillator evolution. The plotted quantity is the absolute deviation $\Delta$ as a function of the number of measurements $N$. Different marker styles and line types correspond to different coherent-state amplitudes $\alpha$ and truncation thresholds $\epsilon$.}
\label{fig:zeno_error}
\end{figure}

In all numerical results presented in the main text [Fig.~1], the Fock-space cutoff was fixed at $n_{\max}=100$, which was verified to be sufficiently large to ensure convergence of all relevant observables and probabilities within the required numerical precision. To further quantify the numerical accuracy, Table~\ref{tab:zeno_error} and Fig.~\ref{fig:zeno_error} compare the numerically evaluated survival probability after $N$ intermediate measurements, $P_{0}^{\mathrm{with,num}}$, with the corresponding analytical expression, $P_{0}^{\mathrm{with,ana}}$. For different values of $\alpha$ and target precision $\varepsilon$, we determine the minimum cutoff dimension $n_{\max}$ for which the deviation
$\Delta=\left|P_{0}^{\mathrm{with,num}}-P_{0}^{\mathrm{with,ana}}\right|$
remains sufficiently small over $N$ measurement rounds.
\medskip

\section{Sec.~III: Alternative implementation of the complete measurement instrument}\label{SecIII}

In the main text, we discuss an optomechanical implementation of sequential measurements that is sufficient for evaluating the bounds $U_{\rm NDC}$ and $L_{\rm NDC}$. However, determining the exact value of $\kappa_{\rm NDC}$ requires implementing the complete quantum instrument associated with the measurement process. In this section, we briefly summarize the optomechanical protocol and derive the corresponding effective quantum instrument it realizes.
 

\subsection{Total quantum instrument implementation}

We now show that the double-swap protocol realizes the measurement instrument mentioned in the footnote of the main text. 

Let the optical ancilla is initialized in the vacuum state $|0\rangle_o$, and the mechanical oscillator is initially in an arbitrary coherent state $|\beta\rangle_m$. Hence, the initial joint state is
\begin{equation*}
\rho_{\mathrm{in}}
=
|0\rangle_o\langle0|
\otimes
|\beta\rangle_m\langle\beta|.
\label{eq:rho_in}
\end{equation*}

\textit{Step 1: First swap.} A beam-splitter interaction implementing a perfect swap gives
\begin{equation*}
\rho_1
=
|\beta\rangle_o\langle\beta|
\otimes
|0\rangle_m\langle0|.
\label{eq:rho1}
\end{equation*}

\textit{Step 2: Optical displacement.}

Applying the displacement operator
$$
D_o(-\alpha)
$$
yields
\begin{equation*}
\rho_2
=
|\beta-\alpha\rangle_o\langle\beta-\alpha|
\otimes
|0\rangle_m\langle0|.
\label{eq:rho2}
\end{equation*}

\textit{Step 3: On/off detection.}

The optical mode is measured using the POVM
\begin{equation*}
E_0
=
|0\rangle_o\langle0|,
\qquad
E_1
=
I_o-|0\rangle_o\langle0|.
\label{eq:povm}
\end{equation*}

For the no-click outcome,
\begin{equation*}
\rho_{2,\mathrm{no}}
=
e^{-|\beta-\alpha|^2}
|0\rangle_o\langle0|
\otimes
|0\rangle_m\langle0|.
\label{eq:rho_no}
\end{equation*}

For the click outcome,
\begin{equation*}
\rho_{2,\mathrm{click}}
=
|\psi_{\mathrm{click}}^{(o)}\rangle
\langle\psi_{\mathrm{click}}^{(o)}|
\otimes
|0\rangle_m\langle0|,
\label{eq:rho_click}
\end{equation*}
where
\begin{equation*}
|\psi_{\mathrm{click}}^{(o)}\rangle
=
|\beta-\alpha\rangle
-
e^{-|\beta-\alpha|^2/2}|0\rangle .
\label{eq:click_state}
\end{equation*}

\textit{Step 4: Second swap.}

After applying the inverse displacement and performing the second swap, the mechanical output states become
\begin{align*}
\rho_{m,\mathrm{no}}
&=
e^{-|\beta-\alpha|^2}
|\alpha\rangle_m\langle\alpha|,
\nonumber\\[4pt]
\rho_{m,\mathrm{click}}
&=
(I_m-|\alpha\rangle\langle\alpha|)
\rho_m
(I_m-|\alpha\rangle\langle\alpha|).
\label{eq:mechanical_outputs}
\end{align*}

The corresponding Kraus operators are therefore
\begin{equation*}
K_{\mathrm{no}}
=
|\alpha\rangle\langle\alpha|,
\qquad
K_{\mathrm{click}}
=
I_m-|\alpha\rangle\langle\alpha|.
\label{eq:kraus}
\end{equation*}

Finally,
\begin{equation*}
K_{\mathrm{no}}^\dagger K_{\mathrm{no}}
+
K_{\mathrm{click}}^\dagger K_{\mathrm{click}}
=
I_m,
\label{eq:completeness}
\end{equation*}
confirming that the protocol realizes a valid quantum measurement instrument~\hyperlink{1}{\textcolor{blue}{[1]}}.

\section{Sec. IV: Detailed calculations for a realistic experimental implementation under environmental damping}\label{SecIV}
In the previous Sec.~III, the measurement together with the associated
outcome-conditioned state updates therefore define a valid
quantum measurement instrument. While the exact
finite-$N$ evaluation of $\kappa_{\mathrm{NDC}}$ requires all
measurement records generated by both $K_{\mathrm{no}}$ and
$K_{\mathrm{click}}$, the analytical lower bound follows from the
single trajectory where every measurement gives the no-click
outcome. Therefore, the dissipative analysis below only requires
the branch $K_{\mathrm{no}}=|\alpha\rangle\langle\alpha|$.

In this implementation, one measurement cycle consists of a
free mechanical evolution of duration $\delta t_{\mathrm{free}}$
followed by one optomechanical swap of duration
$t_{\mathrm{swap}}$ and optical readout. Hence,
$\delta t=t_{\mathrm{swap}}+\delta t_{\mathrm{free}}$ and
$t_f=N\delta t$, with coherent operation requiring
$t_f\ll\Gamma_m^{-1}$. In this section, we now evaluate the effect of damping
on this no-click trajectory.
\subsection{Free mechanical evolution}

During the free evolution interval, only the mechanical mode interacts with its thermal environment. The reduced mechanical density operator obeys the Lindblad master equation~\hyperlink{2}{\textcolor{blue}{[2]}}
\begin{equation*}
\dot{\rho}_m
=
-\frac{i}{\hbar}[H_m,\rho_m]
+
\Gamma_m(\bar n_m^{b}+1)\mathcal{D}[b]\rho_m
+
\Gamma_m\bar n_m^{b}\mathcal{D}[b^\dagger]\rho_m ,
\label{eq:master_free}
\end{equation*}
where $H_m$ is the mechanical Hamiltonian defined in Eq.~\eqref{eq:harmonic_hamiltonian},
$$
\mathcal{D}[O]\rho
=
O\rho O^\dagger
-
\frac{1}{2}
\left\{
O^\dagger O,
\rho
\right\}
$$
denotes the standard Lindblad dissipator, and
$$
\bar n_m^{b}
=
\left(
e^{\hbar\Omega_m/(k_B T_b)}-1
\right)^{-1}
$$
is the mean thermal occupation number of the mechanical reservoir.

For an initial coherent state
$$
|\alpha\rangle_m,
$$
the state remains Gaussian throughout the evolution. After a free evolution time $\delta t$, the mechanical coherent amplitude evolves as
\begin{equation*}
\alpha
\rightarrow
\beta_0
=
\alpha
\exp
\left[
-
\left(
\frac{\Gamma_m}{2}
+
i\Omega_m
\right)\delta t
\right].
\label{eq:beta0}
\end{equation*}

At finite temperature, damping additionally produces Gaussian thermal broadening proportional to $\bar n_m^b$. In the low-temperature limit
$$
\bar n_m^b\rightarrow0,
$$
thermal diffusion vanishes and the mechanical state remains pure:
\begin{equation*}
\rho_m(\delta t)
=
|\beta_0\rangle_m\langle\beta_0|.
\label{eq:rho_free_lowT}
\end{equation*}

At the end of the free evolution interval, the optical mode is initialized in a coherent state
$$
|\alpha_0\rangle_o.
$$
The initial state entering the optomechanical interaction is therefore
\begin{equation*}
\rho(\delta t)
=
|\alpha_0\rangle_o\langle\alpha_0|
\otimes
|\beta_0\rangle_m\langle\beta_0|.
\label{eq:rho_initial_swap}
\end{equation*}

\subsection{Optomechanical swap dynamics}

During the optomechanical interaction, both optical and mechanical modes couple to independent thermal reservoirs. The full two-mode density operator obeys~\hyperlink{2}{\textcolor{blue}{[2]}}
\begin{align}
\dot{\rho}
&=
-\frac{i}{\hbar}[H,\rho]
+
\kappa_o(\bar n_o^b+1)\mathcal{D}[a]\rho
+
\kappa_o\bar n_o^b\mathcal{D}[a^\dagger]\rho
+
\Gamma_m(\bar n_m^b+1)\mathcal{D}[b]\rho
+
\Gamma_m\bar n_m^b\mathcal{D}[b^\dagger]\rho .
\label{eq:master_two_mode}
\end{align}

For a laser--cavity detuning $\Delta=\omega_L-\omega_c=-\Omega_m$ (red detuning), where $\omega_L$ denotes the driving laser frequency and $\omega_c$ denotes the cavity resonance frequency. In the resolved-sideband regime $\Omega_m\gg\kappa_o$, and within the rotating-wave approximation ($g\ll\Omega_m$), the interaction reduces to the beam-splitter Hamiltonian~\cite{Aspelmeyer2014},
\begin{equation*}
H=\hbar\Omega_m(a^\dagger a+b^\dagger b)
-\hbar g(a^\dagger b\,e^{-i\pi/2}+ab^\dagger e^{i\pi/2}).
\label{eq:H}
\end{equation*}

Using the Glauber-Sudarshan $P$ representation, the two-mode density operator is expanded as
\begin{equation}
\rho(t)
=
\int d^2\alpha\,d^2\beta\;
P(\alpha,\beta,t)\,
|\alpha,\beta\rangle
\langle\alpha,\beta|,
\label{eq:P_representation}
\end{equation}
where
$$
|\alpha,\beta\rangle
=
|\alpha\rangle_o\otimes|\beta\rangle_m
$$
denotes the tensor-product coherent state of the optical and mechanical modes.

To convert the master equation into a phase-space equation, we use the standard coherent-state differential identities~\hyperlink{2}{\textcolor{blue}{[2]}}
\begin{align}
a|\alpha\rangle\langle\alpha|
&=
\alpha
|\alpha\rangle\langle\alpha|,
\nonumber\\
a^\dagger|\alpha\rangle\langle\alpha|
&=
\left(
\partial_\alpha+\alpha^*
\right)
|\alpha\rangle\langle\alpha|,
\nonumber\\
|\alpha\rangle\langle\alpha|a^\dagger
&=
\alpha^*
|\alpha\rangle\langle\alpha|,
\nonumber\\
|\alpha\rangle\langle\alpha|a
&=
\left(
\partial_{\alpha^*}+\alpha
\right)
|\alpha\rangle\langle\alpha|,
\label{eq:coherent_state_identities}
\end{align}
together with identical relations for the mechanical operators $b$ and $b^\dagger$.

The interaction Hamiltonian used in the main text is
\begin{equation*}
H_{\mathrm{int}}
=
-\hbar g
\left(
a^\dagger b\,e^{-i\pi/2}
+
ab^\dagger e^{i\pi/2}
\right).
\label{eq:interaction_hamiltonian}
\end{equation*}

Using
$e^{-i\pi/2}=-i,\qquad e^{i\pi/2}=i,$ the interaction Hamiltonian may equivalently be written as
\begin{equation*}
H_{\mathrm{int}}
=
i\hbar g
(a^\dagger b-ab^\dagger).
\label{eq:interaction_hamiltonian_phase}
\end{equation*}

Substituting Eq.~\eqref{eq:P_representation} into Eq.~\eqref{eq:master_two_mode}, the coherent amplitudes $\alpha$ and $\beta$ become phase-space variables, while the creation and annihilation operators are converted into differential operators using Eq.~\eqref{eq:coherent_state_identities}. For the coherent interaction term, we obtain
\begin{align*}
-\frac{i}{\hbar}[H_{\mathrm{int}},\rho]
&=
g
[a^\dagger b-ab^\dagger,\rho]
\nonumber\\
&\rightarrow
-
\partial_\alpha
(-ig\beta P)
-
\partial_\beta
(-ig\alpha P)
-
\partial_{\alpha^*}
(ig\beta^*P)
-
\partial_{\beta^*}
(ig\alpha^*P).
\end{align*}

Similarly, the dissipative Lindblad terms generate both drift and diffusion contributions. After integrating by parts, the master equation reduces to the complex Fokker-Planck equation
\begin{align}
\frac{\partial P}{\partial t}
&=
-
\partial_\alpha
\left(
A_\alpha P
\right)
-
\partial_{\alpha^*}
\left(
A_{\alpha^*}P
\right)
-
\partial_\beta
\left(
A_\beta P
\right)
-
\partial_{\beta^*}
\left(
A_{\beta^*}P
\right)
+
\kappa_o\bar n_o^b
\,
\partial_\alpha\partial_{\alpha^*}P
+
\Gamma_m\bar n_m^b
\,
\partial_\beta\partial_{\beta^*}P,
\label{eq:FP_complex}
\end{align}
where the drift functions are
\begin{align}
A_\alpha
&=
-
\left(
\frac{\kappa_o}{2}
+
i\Omega_m
\right)\alpha
-
ig\beta,
\nonumber\\
A_\beta
&=
-
\left(
\frac{\Gamma_m}{2}
+
i\Omega_m
\right)\beta
-
ig\alpha .
\label{eq:drift_functions}
\end{align}

The terms proportional to $\Omega_m$ describe the fast harmonic rotation of the coherent amplitudes. To isolate the slower dissipative and swap dynamics, we move to the rotating frame
\begin{equation}
\alpha=e^{-i\Omega_m t}\tilde\alpha,
\qquad
\beta=e^{-i\Omega_m t}\tilde\beta.
\label{eq:rotating_frame}
\end{equation}

Using
\begin{align}
\dot\alpha
&=
e^{-i\Omega_m t}
\left(
\dot{\tilde\alpha}
-i\Omega_m\tilde\alpha
\right),
\nonumber\\
\dot\beta
&=
e^{-i\Omega_m t}
\left(
\dot{\tilde\beta}
-i\Omega_m\tilde\beta
\right),
\label{eq:rotating_derivatives}
\end{align}
the explicit oscillatory terms proportional to $\Omega_m$ cancel, yielding
\begin{align}
\dot{\tilde\alpha}
&=
-
\frac{\kappa_o}{2}\tilde\alpha
-
ig\tilde\beta,
\nonumber\\
\dot{\tilde\beta}
&=
-
\frac{\Gamma_m}{2}\tilde\beta
-
ig\tilde\alpha .
\label{eq:rotating_drift}
\end{align}

Introducing real quadratures
\begin{equation}
\tilde\alpha=x+iy,
\qquad
\tilde\beta=u+iv,
\label{eq:real_quadratures}
\end{equation}
and separating Eq.~\eqref{eq:rotating_drift} into real and imaginary parts gives
\begin{align}
\dot x
&=
-
\frac{\kappa_o}{2}x
+
gv,
\nonumber\\
\dot y
&=
-
\frac{\kappa_o}{2}y
-
gu,
\nonumber\\
\dot u
&=
-
\frac{\Gamma_m}{2}u
+
gy,
\nonumber\\
\dot v
&=
-
\frac{\Gamma_m}{2}v
-
gx.
\label{eq:real_drift_equations}
\end{align}

Defining the phase-space vector
\begin{equation*}
\mathbf X
=
(x,y,u,v)^T,
\label{eq:phase_space_vector}
\end{equation*}
Eq.~\eqref{eq:real_drift_equations} can be written compactly as
\begin{equation}
\dot{\mathbf X}
=
A\mathbf X+\xi(t),
\label{eq:OU_process}
\end{equation}
where $\xi(t)$ denotes Gaussian thermal noise and the drift matrix is
\begin{equation}
A=
\begin{pmatrix}
 -\frac{\kappa_o}{2} & 0 & 0 & g \\
 0 & -\frac{\kappa_o}{2} & -g & 0 \\
 0 & g & -\frac{\Gamma_m}{2} & 0 \\
 -g & 0 & 0 & -\frac{\Gamma_m}{2}
\end{pmatrix}.
\label{eq:drift_matrix}
\end{equation}

The diffusion matrix follows directly from the thermal diffusion terms in Eq.~\eqref{eq:FP_complex}:
\begin{equation}
D
=
\frac12
\begin{pmatrix}
\kappa_o\bar n_o^b & 0 & 0 & 0\\
0 & \kappa_o\bar n_o^b & 0 & 0\\
0 & 0 & \Gamma_m\bar n_m^b & 0\\
0 & 0 & 0 & \Gamma_m\bar n_m^b
\end{pmatrix}.
\label{eq:diffusion_matrix}
\end{equation}

The formal solution of Eq.~\eqref{eq:OU_process} is
\begin{equation*}
\mathbf X(t)
=
E(t)\mathbf X(0)
+
\int_0^t ds\;
E(t-s)\xi(s),
\label{eq:formal_solution}
\end{equation*}
where the propagator is
\begin{equation*}
E(t)=e^{At}.
\label{eq:formal_propagator}
\end{equation*}

To evaluate the propagator analytically, we consider the strong-coupling regime
\begin{equation*}
g\gg\kappa_o,\Gamma_m .
\label{eq:strong_coupling}
\end{equation*}

Defining
\begin{equation}
\gamma
=
\frac{\kappa_o+\Gamma_m}{4},
\qquad
\delta
=
\frac{\Gamma_m-\kappa_o}{4},
\label{eq:gamma_delta}
\end{equation}
the eigenvalues of the drift matrix become
\begin{equation*}
\lambda_\pm
=
-\gamma
\pm
\sqrt{\delta^2-g^2}.
\label{eq:eigenvalues_exact}
\end{equation*}

Since
$$
g\gg|\delta|,
$$
the square root simplifies to
\begin{equation}
\sqrt{\delta^2-g^2}
\simeq
ig,
\label{eq:strong_coupling_approx}
\end{equation}
and therefore
\begin{equation*}
\lambda_\pm
\simeq
-\gamma
\pm
ig.
\label{eq:eigenvalues_sc}
\end{equation*}

Thus, the dynamics consists of coherent beam-splitter oscillations at frequency $g$ together with exponential damping at rate $\gamma$. Consequently, the propagator reduces to
\begin{equation}
E(t)
\simeq
e^{-\gamma t}R(t),
\label{eq:propagator}
\end{equation}
where
\begin{equation}
R(t)=
\begin{pmatrix}
\cos(gt) & 0 & 0 & \sin(gt) \\
0 & \cos(gt) & -\sin(gt) & 0 \\
0 & \sin(gt) & \cos(gt) & 0 \\
-\sin(gt) & 0 & 0 & \cos(gt)
\end{pmatrix}.
\label{eq:rotation_matrix}
\end{equation}

The covariance matrix follows from the standard Ornstein-Uhlenbeck solution
\begin{equation}
V(t)
=
\int_0^t ds\;
E(s)DE^T(s).
\label{eq:covariance_general}
\end{equation}

Substituting Eqs.~\eqref{eq:propagator} and \eqref{eq:diffusion_matrix} into Eq.~\eqref{eq:covariance_general}, the optical quadrature variance becomes
\begin{align}
V_{xx}(t)
&=
\int_0^t ds
\left[
D_{11}E_{11}^2(s)
+
D_{44}E_{14}^2(s)
\right]
\nonumber\\
&=
\frac12
\int_0^t ds\;
e^{-2\gamma s}
\left[
\kappa_o\bar n_o^b\cos^2(gs)
+
\Gamma_m\bar n_m^b\sin^2(gs)
\right].
\label{eq:variance_xx}
\end{align}

Using
\begin{align}
\cos^2(gs)
&=
\frac12
\left[
1+\cos(2gs)
\right],
\nonumber\\
\sin^2(gs)
&=
\frac12
\left[
1-\cos(2gs)
\right],
\label{eq:trig_identities}
\end{align}
Eq.~\eqref{eq:variance_xx} becomes
\begin{align*}
V_{xx}(t)
&=
\frac14
\int_0^t ds\;
e^{-2\gamma s}
\left[
\kappa_o\bar n_o^b
+
\Gamma_m\bar n_m^b
\right]
\nonumber\\
&\quad
+
\frac14
\int_0^t ds\;
e^{-2\gamma s}
\left[
\kappa_o\bar n_o^b
-
\Gamma_m\bar n_m^b
\right]
\cos(2gs).
\label{eq:variance_split}
\end{align*}

In the strong-coupling regime, the oscillatory integral averages out to leading order in $\gamma/g$, yielding
\begin{align*}
V_{xx}(t)
&\simeq
\frac14
\left(
\kappa_o\bar n_o^b
+
\Gamma_m\bar n_m^b
\right)
\int_0^t ds\;
e^{-2\gamma s}.
\end{align*}

Evaluating the integral,
\begin{equation*}
\int_0^t ds\;
e^{-2\gamma s}
=
\frac{1-e^{-2\gamma t}}{2\gamma},
\label{eq:exp_integral}
\end{equation*}
and using
$$
2\gamma
=
\frac{\kappa_o+\Gamma_m}{2},
$$
gives
\begin{equation}
V_{xx}(t)
=
\frac{
\kappa_o\bar n_o^b
+
\Gamma_m\bar n_m^b
}
{2(\kappa_o+\Gamma_m)}
\left[
1-e^{-(\kappa_o+\Gamma_m)t/2}
\right].
\label{eq:variance_xx_final}
\end{equation}
An identical calculation yields
\begin{equation}
V_{yy}(t)
=
V_{uu}(t)
=
V_{vv}(t)
=
V_{xx}(t),
\label{eq:equal_variances}
\end{equation}
up to corrections of order $\gamma/g$.

Since
$$
|\alpha|^2=x^2+y^2,
$$
the complex variances associated with the optical and mechanical coherent amplitudes become
\begin{equation}
V_o(t)
=
V_m(t)
=
2V_{xx}(t).
\label{eq:complex_variance}
\end{equation}

Substituting Eq.~\eqref{eq:variance_xx_final} into Eq.~\eqref{eq:complex_variance}, we obtain
\begin{equation}
V_o(t)
=
V_m(t)
=
\frac{
\kappa_o\bar n_o^b
+
\Gamma_m\bar n_m^b
}{
\kappa_o+\Gamma_m
}
\left[
1-e^{-(\kappa_o+\Gamma_m)t/2}
\right].
\label{eq:variance}
\end{equation}

The optomechanical state swap occurs at
\begin{equation}
t_{\mathrm{swap}}
=
\frac{\pi}{2g}.
\label{eq:swap_time}
\end{equation}

Substituting Eq.~\eqref{eq:swap_time} into Eq.~\eqref{eq:variance}, the effective variance at the swap time becomes
\begin{equation}
V_{\mathrm{swap}}
=
\frac{
\kappa_o\bar n_o^b
+
\Gamma_m\bar n_m^b
}{
\kappa_o+\Gamma_m
}
\left[
1-e^{-(\kappa_o+\Gamma_m)\pi/(4g)}
\right].
\label{eq:variance_swap}
\end{equation}

The corresponding Green function therefore remains Gaussian throughout the evolution. At the swap time $t_{\mathrm{swap}}$, it factorizes as
\begin{equation*}
G(\alpha,\beta,t_{\mathrm{swap}})
=
G_o(\alpha|t_{\mathrm{swap}})\,
G_m(\beta|t_{\mathrm{swap}}),
\label{eq:green_factorized}
\end{equation*}
where
\begin{equation*}
G_o(\alpha|t_{\mathrm{swap}})
=
\frac{1}{\pi V_{\mathrm{swap}}}
\exp
\left[
-
\frac{
|\alpha-\alpha_{\mathrm{cl}}(t_{\mathrm{swap}})|^2
}{
V_{\mathrm{swap}}
}
\right],
\label{eq:green_optical}
\end{equation*}
and similarly for the mechanical mode,
\begin{equation*}
G_m(\beta|t_{\mathrm{swap}})
=
\frac{1}{\pi V_{\mathrm{swap}}}
\exp
\left[
-
\frac{
|\beta-\beta_{\mathrm{cl}}(t_{\mathrm{swap}})|^2
}{
V_{\mathrm{swap}}
}
\right].
\label{eq:green_mechanical}
\end{equation*}

The two-mode density operator at the swap time is therefore
\begin{align*}
\rho(t_{\mathrm{swap}})
&=
\int d^2\alpha\,d^2\beta\;
G(\alpha,\beta,t_{\mathrm{swap}})\,
|\alpha,\beta\rangle
\langle\alpha,\beta|
\nonumber\\
&=
\rho_o(t_{\mathrm{swap}})\otimes\rho_m(t_{\mathrm{swap}}),
\label{eq:factorized}
\end{align*}
with
\begin{align*}
\rho_o(t_{\mathrm{swap}})
&=
\int d^2\alpha\;
G_o(\alpha|t_{\mathrm{swap}})\,
|\alpha\rangle\langle\alpha|,
\nonumber\\
\rho_m(t_{\mathrm{swap}})
&=
\int d^2\beta\;
G_m(\beta|t_{\mathrm{swap}})\,
|\beta\rangle\langle\beta|.
\end{align*}

In the low-temperature limit
$$
\bar n_o^b,\bar n_m^b\rightarrow0,
$$
the variance $V_{\mathrm{swap}}$ vanishes, and the Gaussian distributions collapse into delta functions:
\begin{align*}
G_o(\alpha|t_{\mathrm{swap}})
&\rightarrow
\delta^{(2)}
\left(
\alpha-\alpha_{\mathrm{cl}}(t_{\mathrm{swap}})
\right),
\nonumber\\
G_m(\beta|t_{\mathrm{swap}})
&\rightarrow
\delta^{(2)}
\left(
\beta-\beta_{\mathrm{cl}}(t_{\mathrm{swap}})
\right).
\end{align*}

Consequently, the final state becomes pure,
\begin{equation*}
\rho(t_{\mathrm{swap}})
=
|\tilde\beta\rangle_o\langle\tilde\beta|
\otimes
|\tilde\alpha\rangle_m\langle\tilde\alpha|.
\label{eq:final_pure}
\end{equation*}

The corresponding coherent amplitudes are obtained from the classical evolution generated by Eq.~\eqref{eq:propagator}. At the swap time,
$$
gt_{\mathrm{swap}}=\frac{\pi}{2},
$$
and therefore
\begin{equation*}
R(t_{\mathrm{swap}})=
\begin{pmatrix}
0 & 0 & 0 & 1 \\
0 & 0 & -1 & 0 \\
0 & 1 & 0 & 0 \\
-1 & 0 & 0 & 0
\end{pmatrix}.
\label{eq:rotation_swap}
\end{equation*}

Thus, the optical and mechanical amplitudes are exchanged under the beam-splitter dynamics, while dissipation contributes the overall damping factor
$$
e^{-\gamma t_{\mathrm{swap}}}.
$$

Restoring the laboratory-frame oscillation from Eq.~\eqref{eq:rotating_frame}, the coherent amplitudes evolve as
\begin{align}
\tilde\beta
&=
e^{-\gamma t_{\mathrm{swap}}}
e^{-i\Omega_m t_{\mathrm{swap}}}
\beta_0,
\nonumber\\[4pt]
\tilde\alpha
&=
e^{-\gamma t_{\mathrm{swap}}}
e^{-i\Omega_m t_{\mathrm{swap}}}
\alpha_0 .
\label{eq:amplitude_intermediate}
\end{align}

Substituting
$$
t_{\mathrm{swap}}=\frac{\pi}{2g},
\qquad
\gamma=\frac{\kappa_o+\Gamma_m}{4},
$$
gives
\begin{align*}
\tilde\beta
&=
\exp
\left[
-
\frac{\pi(\kappa_o+\Gamma_m)}{8g}
\right]
\exp
\left[
-
i\Omega_m\frac{\pi}{2g}
\right]
\beta_0,
\nonumber\\[6pt]
\tilde\alpha
&=
\exp
\left[
-
\frac{\pi(\kappa_o+\Gamma_m)}{8g}
\right]
\exp
\left[
-
i\Omega_m\frac{\pi}{2g}
\right]
\alpha_0 .
\label{eq:final_amplitudes}
\end{align*}

Thus, in the absence of thermal noise, the optomechanical interaction implements an almost ideal coherent-state swap, while dissipation produces only an overall exponential attenuation of the coherent amplitudes.

\subsection{Damped evolution and computation of damped $\kappa_{\mathrm{NDC}}$}

Let $\beta_j$ denote the mechanical coherent amplitude at the beginning of the $j$-th interrogation round, with initial condition
$$
\beta_0=\alpha.
$$
The total protocol duration $t_f$ is divided into $N$ identical cycles according to
\begin{equation}
\frac{t_f}{N}
=
\delta t_{\mathrm{free}}
+
t_{\mathrm{swap}},
\label{eq:cycle_time}
\end{equation}
where each cycle consists of a free mechanical evolution followed by a dissipative optomechanical swap interaction.

During the free-evolution interval $\delta t_{\mathrm{free}}$, the mechanical mode evolves under damping and harmonic rotation, while the optical mode remains in vacuum. For an initial state
$$
|0\rangle_o\otimes|\beta_j\rangle_m,
$$
the evolution gives
\begin{equation*}
|0\rangle_o\otimes|\beta_j\rangle_m
\longrightarrow
|0\rangle_o\otimes|\beta_j B\rangle_m,
\label{eq:free_evolution}
\end{equation*}
where
\begin{equation*}
B
=
\exp
\left[
-
\left(
\frac{\Gamma_m}{2}
+
i\Omega_m
\right)
\delta t_{\mathrm{free}}
\right]
\label{eq:B_definition}
\end{equation*}
is the complex attenuation factor associated with mechanical damping and phase rotation. Consequently, the coherent amplitude transforms as
\begin{equation*}
\beta_j
\longrightarrow
\beta_j B.
\label{eq:beta_free}
\end{equation*}

At time
$$
t=(j+1)\delta t_{\mathrm{free}}+j\,t_{\mathrm{swap}},
$$
an optical ancilla prepared in the coherent state $|\alpha\rangle_o$ interacts with the mechanical mode for a duration
\begin{equation*}
t_{\mathrm{swap}}
=
\frac{\pi}{2g}.
\label{eq:swap_duration}
\end{equation*}

In the strong-coupling regime
$$
g\gg\kappa_o,\Gamma_m,
$$
the optomechanical interaction implements an approximate dissipative beam-splitter swap. The joint coherent state evolves as
\begin{equation*}
|\alpha\rangle_o\otimes|\beta_j B\rangle_m
\longrightarrow
|A\beta_j B\rangle_o
\otimes
|A\alpha\rangle_m,
\label{eq:swap_evolution}
\end{equation*}
where
\begin{equation*}
A=f_1 f_2,
\label{eq:A_definition}
\end{equation*}
with
\begin{align*}
f_1
&=
\exp
\left[
-
\frac{\kappa_o+\Gamma_m}{4}
t_{\mathrm{swap}}
\right],
\nonumber\\
f_2
&=
\exp
\left(
-i\Omega_m t_{\mathrm{swap}}
\right).
\label{eq:f_gamma}
\end{align*}

Thus, the swap interaction exchanges the coherent amplitudes while introducing both damping and phase rotation.

After the swap, the optical mode is measured, while the mechanical mode remains in the coherent state
$$
|A\alpha\rangle_m,
$$
which is independent of the incoming mechanical amplitude $\beta_j$. Therefore, after the first interrogation cycle the dynamics reaches a fixed point,
\begin{equation}
\beta_{j+1}=A\alpha,
\qquad
\beta_j=A\alpha
\quad
(j\ge1).
\label{eq:fixed_point}
\end{equation}


\subsection*{Detection Probabilities}

Immediately after the swap interaction, the optical mode is in the coherent state
$$
|A\beta_j B\rangle_o.
$$
Applying the displacement operator
$$
D_o(-\alpha)
$$
transforms the optical state as
\begin{equation*}
|A\beta_j B\rangle_o
\longrightarrow
|A\beta_j B-\alpha\rangle_o.
\label{eq:displaced_state}
\end{equation*}

A click-no-click measurement described by the POVM
$$
\{
|0\rangle\langle0|,
\,
\mathbb I-|0\rangle\langle0|
\}
$$
then gives the no-click probability
\begin{equation}
P(a_j=0)
=
\exp
\left[
-
|A\beta_j B-\alpha|^2
\right],
\label{eq:no_click_general}
\end{equation}
and the click probability
\begin{equation*}
P(a_j=1)
=
1-P(a_j=0).
\label{eq:click_general}
\end{equation*}

Using the fixed-point relation in Eq.~\eqref{eq:fixed_point}, we obtain
\begin{align*}
P(a_0=0)
&=
\exp
\left[
-
|A\alpha B-\alpha|^2
\right],
\nonumber\\
P(a_j=0)
&=
\exp
\left[
-
|A^2\alpha B-\alpha|^2
\right],
\qquad
(j\ge1).
\label{eq:no_click_explicit}
\end{align*}


\subsection*{Sequential No-Click Probability}

The probability of obtaining no-click outcomes under damping in all $N$ rounds is therefore,
\begin{equation}
\begin{aligned}
   P_{0,\rm damped}^{\text{(with)}}(t_f)
&= P(a_1=0,a_2=0,\ldots,a_{N-1}=0,a_N=0).
\nonumber\\
&=
\exp
\left[
-
|\alpha|^2
\left(
|AB-1|^2
+
(N-1)|A^2B-1|^2
\right)
\right]. 
\end{aligned}
\label{eq:P0_product}
\end{equation}

We now define
\begin{align}
AB
&=
R_1 e^{-i\Theta_1},
\nonumber\\
A^2B
&=
R_2 e^{-i\Theta_2},
\label{eq:RTheta_definitions}
\end{align}
where the magnitudes are
\begin{align*}
R_1
&=
\exp
\left[
-
\frac{\kappa_o+\Gamma_m}{4}t_{\mathrm{swap}}
-
\frac{\Gamma_m}{2}\delta t_{\mathrm{free}}
\right],
\nonumber\\
R_2
&=
\exp
\left[
-
\frac{\kappa_o+\Gamma_m}{2}t_{\mathrm{swap}}
-
\frac{\Gamma_m}{2}\delta t_{\mathrm{free}}
\right],
\label{eq:R_definitions}
\end{align*}
while the corresponding phases are
\begin{align*}
\Theta_1
&=
\Omega_m
\left(
\delta t_{\mathrm{free}}
+
t_{\mathrm{swap}}
\right)
=
\Omega_m\frac{t_f}{N},
\nonumber\\
\Theta_2
&=
\Omega_m
\left(
\delta t_{\mathrm{free}}
+
2t_{\mathrm{swap}}
\right)
=
\Omega_m\frac{t_f}{N}
+
\Omega_m t_{\mathrm{swap}}.
\end{align*}

Using the identity
\begin{equation}
|z-1|^2
=
(1-R)^2
+
2R(1-\cos\Theta),
\qquad
z=Re^{i\Theta},
\label{eq:complex_identity}
\end{equation}
Eq.~\eqref{eq:P0_product} becomes
\begin{align*}
P_{0,\rm damped}^{\text{(with)}}(t_f)
=
\exp
\Big[
-|\alpha|^2
\big(
&
(1-R_1)^2
+
(N-1)(1-R_2)^2
+
2R_1(1-\cos\Theta_1)
\nonumber\\
&
+
2(N-1)R_2(1-\cos\Theta_2)
\big)
\Big].
\label{eq:P0_expanded}
\end{align*}

If the swap duration satisfies
\begin{equation*}
\Omega_m t_{\mathrm{swap}}
=
2\pi q,
\label{eq:swap_condition}
\end{equation*}
then
$$
\Theta_2=\Theta_1+2\pi q,
$$
and therefore
$$
\cos\Theta_2=\cos\Theta_1.
$$
Defining
\begin{equation}
\Theta\equiv\Theta_1,
\label{eq:Theta_definition}
\end{equation}
the sequential no-click probability reduces to
\begin{align}
P_{0,\rm damped}^{\text{(with)}}(t_f)
=
\exp
\Big[
-|\alpha|^2
\big(
&
(1-R_1)^2
+
(N-1)(1-R_2)^2
\nonumber\\
&
+
2(R_1+(N-1)R_2)
(1-\cos\Theta)
\big)
\Big].
\label{eq:P0_simplified}
\end{align}

Next, we choose the total evolution time such that
\begin{equation*}
\Omega_m t_f
=
(2k+1)\pi.
\label{eq:total_time_condition}
\end{equation*}
Using Eq.~\eqref{eq:Theta_definition}, this gives
\begin{equation}
\Theta
=
\frac{(2k+1)\pi}{N}.
\label{eq:Theta_k}
\end{equation}

The integer $k$ is constrained by the condition
\begin{equation*}
Nt_{\mathrm{swap}}
<
t_f
=
\frac{(2k+1)\pi}{\Omega_m}
\ll
\frac{1}{\Gamma_m},
\label{eq:k_constraint}
\end{equation*}
which implies
\begin{equation*}
\frac12
\left(
\frac{\Omega_m}{2g}N-1
\right)
\lesssim
k
\lesssim
\frac12
\left(
\frac{\Omega_m}{\pi\Gamma_m}-1
\right).
\label{eq:k_bounds}
\end{equation*}

Imposing the condition
$$
\Omega_m t_{\mathrm{swap}}=2\pi q,
$$
the minimal consistent choice is
$$
k=qN.
$$
Equation~\eqref{eq:Theta_k} then becomes
\begin{equation*}
\Theta
=
2\pi q+\frac{\pi}{N}
\equiv
\frac{\pi}{N}
\quad
(\mathrm{mod}\;2\pi),
\label{eq:Theta_final}
\end{equation*}
so that
\begin{equation*}
\cos\Theta
=
\cos\left(\frac{\pi}{N}\right).
\label{eq:cosTheta}
\end{equation*}

Consequently,
\begin{align}
P_{0,\rm damped}^{\text{(with)}}(t_f)
=
\exp
\Big[
-|\alpha|^2
\big(
&
(1-R_1)^2
+
(N-1)(1-R_2)^2
\nonumber\\
&
+
2(R_1+(N-1)R_2)
\left(
1-\cos\frac{\pi}{N}
\right)
\big)
\Big].
\label{eq:P0_final}
\end{align}

For large $N$, using
\begin{equation}
1-\cos\left(\frac{\pi}{N}\right)
\approx
\frac{\pi^2}{2N^2},
\label{eq:largeN_cos}
\end{equation}
we obtain
\begin{align*}
P_{0,\rm damped}^{\text{(with)}}(t_f)
\approx
\exp
\Big[
-|\alpha|^2
\big(
&
(1-R_1)^2
+
(N-1)(1-R_2)^2
\nonumber\\
&
+
\frac{\pi^2}{N^2}
(R_1+(N-1)R_2)
\big)
\Big].
\label{eq:P0_largeN}
\end{align*}


\subsection*{Single-Shot Measurement}

For comparison, we now consider a single measurement performed only at the final time $t_f$. Defining
\begin{equation*}
t_f'
=
t_f-t_{\mathrm{swap}},
\label{eq:tfprime}
\end{equation*}
the corresponding no-click probability under damping becomes
\begin{equation*}
P_{0, \rm damped}^{\text{(without)}}(t_f)
=
\exp
\left[
-
|\alpha|^2|AB_f'-1|^2
\right],
\label{eq:Pnc_general}
\end{equation*}
where
\begin{equation*}
AB_f'
=
Re^{-i\Theta},
\label{eq:ABf}
\end{equation*}
with
\begin{equation*}
R
=
\exp
\left[
-
\frac{\Gamma_m}{2}t_f'
\right]
\exp
\left[
-
\frac{\kappa_o+\Gamma_m}{4}t_{\mathrm{swap}}
\right].
\label{eq:R_single}
\end{equation*}

Using Eq.~\eqref{eq:complex_identity}, we obtain
\begin{equation*}
P_{0, \rm damped}^{\text{(without)}}(t_f)
=
\exp
\left[
-
|\alpha|^2
\left(
(1-R)^2
+
2R(1-\cos\Theta)
\right)
\right].
\label{eq:Pnc_expanded}
\end{equation*}

This probability is minimized for
\begin{equation*}
\Theta=(2k+1)\pi,
\label{eq:Pnc_min_condition}
\end{equation*}
yielding
\begin{equation*}
P_{0, \rm damped}^{\text{(without, min)}}(t_f)
=
\exp
\left[
-
|\alpha|^2(R+1)^2
\right].
\label{eq:Pnc_min}
\end{equation*}

We minimize $P_{0, \rm damped}^{\text{(without)}}(t_f)$ in order to maximize the witness
\begin{equation*}
\kappa_{\mathrm{NDC}}^{\mathrm{damped}}
\ge
P_{0, \rm damped}^{\text{(with)}}(t_f)
-
P_{0, \rm damped}^{\text{(without)}}(t_f).
\label{eq:NDC_bound}
\end{equation*}


\subsection{Ideal Limit}

In the ideal dissipation-free limit
$$
\kappa_o=\Gamma_m=0,
$$
one has
$$
A=B=e^{-i\Omega_m t},
$$
so that
$$
R_1=R_2=1.
$$
Equation~\eqref{eq:P0_final} then reduces to
\begin{equation*}
P_{0, \rm damped}^{\text{(with)}}(t_f)
=
\exp
\left[
-
2|\alpha|^2
N
\left(
1-\cos\frac{\pi}{N}
\right)
\right].
\label{eq:P0_ideal}
\end{equation*}

For large $N$, using
$$
1-\cos(\pi/N)\approx \frac{\pi^2}{2N^2},
$$
we obtain
\begin{equation*}
P_{0, \rm damped}^{\text{(with)}}(t_f)
\approx
\exp
\left[
-
\frac{\pi^2|\alpha|^2}{N}
\right].
\label{eq:P0_ideal_largeN}
\end{equation*}

Similarly, the single-shot probability becomes
\begin{equation*}
P_{0, \rm damped}^{\text{(without, min)}}(t_f)
=
\exp[-4|\alpha|^2].
\label{eq:Pnc_ideal}
\end{equation*}

\hrule
\vspace{0.5cm}
\noindent \hypertarget{QuantumMeasurement2016}{[1]} P. Busch, P. Lahti, J.-P. Pellonp\"aa, and K. Ylinen.
\textit{Quantum Measurement}.
Springer (2016).

\noindent \hypertarget{Carmichael2013}{[2]} H. Carmichael.
\textit{Statistical Methods in Quantum Optics}.
Springer (2013).\\

\end{document}